\documentclass{ws-ijmpb}
\usepackage{amssymb}

\renewcommand{\theequation}{\arabic{section}.\arabic{equation}}

\def\a{\alpha}
\def\b{\beta}
\def\d{\delta}
\def\e{\epsilon}
\def\g{\gamma}
\def\k{\kappa}
\def\l{\lambda}
\def\m{\mu}

\def\O{\Omega}
\def\P{\Phi}
\def\s{\sigma}
\def\t{\tilde}
\def\x{\xi}

\def\str{\mbox{str}}

\def\Zb{\mathbb{Z}}

\def\GL{U_q(gl(m|n))}
\def\gl{U_q(gl(2|1))}

\def\S{{\cal S}}

\def\beq{\begin{equation}}
\def\eeq{\end{equation}}
\def\bea{\begin{eqnarray}}
\def\eea{\end{eqnarray}}
\def\ba{\begin{array}}
\def\ea{\end{array}}
\def\no{\nonumber}

\def\lt{\left}
\def\rt{\right}

\newtheorem{Theorem}{Theorem}
\newtheorem{Definition}{Definition}
\newtheorem{Proposition}{Proposition}

\makeatletter
\renewcommand\thesection{\@arabic\c@section}
\renewcommand\thesubsection{\thesection.\@arabic\c@subsection}
\makeatother

\newcommand{\sect}[1]{\setcounter{equation}{0}\section{#1}}
\renewcommand{\theequation}{\thesection.\arabic{equation}}

\begin{document}

\markboth{Shao-You Zhao et al} {On the Construction of Correlation
functions}


\title{On the Construction of Correlation Functions for the
Integrable Supersymmetric Fermion Models}

\author{Shao-You Zhao${}^{a}$, Wen-Li Yang${}^{a,b}$, and ~Yao-Zhong
 Zhang${}^a$}

\address{${}^a$ Department of Mathematics, University of Queensland,
            Brisbane, QLD 4072, Australia\\
     ${}^b$ Institute of Modern Physics, Northwest University,
     Xian 710069, P.R. China\\[3mm]
    syz@maths.uq.edu.au; wenli@maths.uq.edu.au; yzz@maths.uq.edu.au\\
    {Int. J. Mod. Phys.} {\bf B20} No.5 (2006) 505-549 (hep-th/0601065)}


\maketitle




\def\R{\overline{R}}



\begin{abstract}
We review the recent progress on the construction of the
determinant representations of the correlation functions for the
integrable supersymmetric fermion models. The factorizing
$F$-matrices (or the so-called $F$-basis) play an important role
in the construction. In the $F$-basis, the creation (and the
annihilation) operators and the Bethe states of the integrable
models are given in completely symmetric forms. This leads to the
determinant representations of the scalar products of the Bethe
states for the models. Based on the scalar products, the
determinant representations of the correlation functions may be
obtained. As an example, in this review, we give the determinant
representations of the two-point correlation function for the
$\gl$ (i.e. $q$-deformed) supersymmetric $t$-$J$ model. The
determinant representations are useful for analysing physical
properties of the integrable models in the thermodynamical limit.

\vspace{1truecm}


\end{abstract}

\newpage
\sect{Introduction}

It is well known that quantum integrable and exactly solvable
systems based on the Yang-Baxter equation (YBE) play an important
role in modern mathematics and physics. They have important
applications in a startling variety of physical theories, such as
the theory of the ultrasmall metallic grains (see e.g.
refs.\cite{Dukelsky04,Delft01} and references therein), the
(${\cal N}=4$) four dimensional super-symmetric Yang-Mills gauge
theories (see e.g. refs.\cite{DHoker99,Minahan02,Kazakov04} and
references therein), and string theories (see e.g.
refs.\cite{Tseytlin03,Beisert03} and references therein).

In dealing with the integrable systems, the algebraic Bethe ansatz
or the Quantum Inverse Scattering Method (QISM) provides a
powerful tool to the diagonalization of their Hamiltonians. In
this approach, Bethe states are constructed by the pseudo-particle
creation operators which are from the off-diagonal entries of the
monodromy matrix. After obtaining the eigenvalues of a system, one
of the most interesting and challenging problems is to construct
scalar products (including the norms) of the eigenstates and
correlation functions \cite{Korepin93,Smirnov92}. In 1981, Gaudin
et al proposed a hypothesis that the norm of the coordinate
eigenstates for the Heisenberg XXZ spin chain model is given by
some Jacobians \cite{Gaudin81}. This hypothesis was proved
completely by Korepin in \cite{Korepin82}. Moreover based on the
results of this work, the authors in
\cite{Izergin84,Izergin85,Korepin84} computed the correlation
functions for the integrable XXX and XXZ models as well as the
one-dimensional Bose Gas system. With the help of the auxiliary
dual quantum fields \cite{Korepin87} and the determinant
representation for the partition function of the six-vertex model
with domain wall boundary conditions \cite{Izergin87,Izergin92},
the determinant representations of the correlation functions of
the XXX and XXZ models were obtained \cite{Korepin93,Essler95}.
Let us remark that under the hypothesis concerning the space of
physical states \cite{Jimbo}, integral representations of
correlation functions for integrable models on an infinite 1-d
lattice can be obtained by using the technique of the $q$-deformed
vertex operator \cite{FR}${}^-$\cite{Yang9902}.

In 1996, Maillet et al \cite{Maillet96} proved that for the
$R$-matrices of the Heisenberg XXX and XXZ spin chain systems,
there exist non-degenerate lower-triangular $F$-matrices (i.e. the
Drinfeld twists) \cite{Drinfeld83} with which the $R$-matrices are
factorized
\begin{eqnarray}
R_{12}(\l_1,\l_2)=F^{-1}_{21}(\l_2,\l_1)F_{12}(\l_1,\l_2),
\label{eq:Drinfeld}
\end{eqnarray}
where $R\in End(V\otimes V)$ with $V$ being the 2-dimensional
$gl(2)$ or $U_q(gl(2))$ module.
%
%
Working in the basis provided by the $N$-site $F$-matrix, i.e. the
so-called $F$-basis, they proved that the entries of the monodromy
matrix and therefore the Bethe states of the systems are
simplified to take completely symmetric forms. This observation
implies that the exact evaluation of scalar products and
correlation functions of a integrable system is feasible within
the framework of the algebraic Bethe ansatz. In \cite{Kitanine98},
Kitanine et al obtained the determinant representation of the
correlation functions for the XXX and XXZ models, and showed the
scalar products and norms of the eigenstates of the systems
obtained using the Drinfeld twist approach coincide with those
obtained in \cite{Gaudin81,Korepin82,Izergin87}.

The Drinfeld twist approach in \cite{Maillet96,Kitanine98} was
generalized to other cases. In \cite{Terras99}, the Drinfeld
twists associated with any finite-dimensional irreducible
representations of the Yangian $Y[gl(2)]$ were investigated, and
in \cite{Kitanine01} the correlation functions for the higher spin
XXX chains were computed.  In \cite{Izergin98}, the spontaneous
magnetization of the XXZ chain on the finite lattice was
represented. In \cite{Albert00}, Albert et al constructed the
$F$-matrix of the $gl(m)$ rational Heisenberg model and obtained a
polarization free representation of the creation operators. Using
these results, they resolved the hierarchy of the nested Bethe
ansatz for the $gl(m)$ model. In \cite{Albert0005,Albert0007}, the
Drinfeld twists of the elliptic XYZ model and Belavin model were
constructed. In \cite{zsy0506}, we obtained the determinant
representations for the $U_q(gl(1|1))$ free fermion model.

As far as we know, the determinant representations of correlation
functions were known only for integrable models related to $gl(2)$
algebra, and it had been a major longstanding problem to find the
determinant representations of correlation functions for
integrable models related to $gl(3)$ or other higher rank
algebras. Very recently in \cite{zsy0511}, based on our results on
the Drinfeld twists and symmetric Bethe states
\cite{zsy04,zsy0502,zsy0503}, we have presented a solution to this
problem for $gl(2|1)$ algebra which is a graded version of
$gl(3)$, and found the determinant representations of the
correlation functions for the supersymmetric $t$-$J$ model.

In this article, we review the recent progress that we made on the
Drinfeld twists and on the determinant representations of the
correlation functions for supersymmetric integrable models such as
the supersymmetric $t$-$J$ model. Supersymmetric integrable models
form an important class of exactly soluble models
\cite{Per81}${}^-$\cite{Kul86} as they provide strongly correlated
fermion systems of superconductivity
\cite{Essler9211}${}^-$\cite{yzz95} and have important
applications in the AdS/CFT correspondence. In section 2, we
briefly review the background of the integrable models and
introduce the $N$-site $R$-matrix. In section 3, we discuss the
properties of the monodromy matrices, and give the recursive
relations as well as the representations of the local generators
in terms of monodromy matrix elements. In section 4, we derive the
factorizing $F$-matrix and its inverse. Then in section 5, we
discuss the monodromy matrices in the $F$-basis. In section 6, we
describe the construction of the determinant representations of
scalar products and correlation functions, using the $q$-deformed
supersymmetric $t$-$J$ model as an example. In section 7, we
conclude the review by offering some discussions and outlooks.

\sect{Integrable $U_q(gl(m|n))$ supersymmetric model} \label{QSA}
\setcounter{equation}{0} We first introduce some useful properties
of the quantum superalgebra $\GL$. For more details, see
refs.\cite{zsy0503,ZhangRB92}. Let us fix two non-negative
integers $n$, $m$ such that $m+n\geq 2$ and a positive integer
$N\,(\geq 2)$, and a generic complex number $\eta$ such that the
q-deformation parameter, related to $\eta$ through $q=e^{\eta}$,
is not a root of unity. Let $V$ be a $\Zb_2$-graded
$(m+n)$-dimensional vector space with the orthonormal basis
$\{|i\rangle,\,i=1,\ldots,m+n\}$. The $\Zb_2$-grading is chosen
as: $[1]=\ldots=[m]=1,\,[m+1]=\ldots=[m+n]=0$.

The quantum superalgebra $\GL$ is a $\Zb_2$-graded unital
associative superalgebra   generated by the Cartan generators
$E^{i,i},\,(i=1,\ldots,m+n)$ and the simple root generators
$E^{j,j+1},\,E^{j+1,j}\,(j=1,\ldots,m+n-1)$  with the
$\Zb_2$-grading $[E^{i,i}]=0$,
$[E^{j+1,j}]=[E^{j,j+1}]=[j]+[j+1]$. The $\Zb_2$-graded vector
space $V$ supplies  the fundamental $\GL$-module
 and the generators of $\GL$ are represented in this space by
$ \pi(E^{i,i})=e_{i,i},~
\pi(E^{j,j+1})=e_{j,j+1},~\pi(E^{j+1,j})=e_{j+1,j},$ where
$e_{i,j}\in{\rm End}(V)$ is the elementary matrix with elements
$(e_{i,j})^l_k=\d_{jk}\d_{il}$.

With the help the simple root generators, we can construct the
non-simple root generators as follows
\begin{eqnarray}
&&E^{\a,\g}=E^{\a,\b}E^{\b,\g}
   -q^{-(-1)^{[\b]}}E^{\b,\g}E^{\a,\b},~~1\leq\a<\b<\g\leq m+n,
   \label{non-simple1}\\
   &&E^{\g,\a}=E^{\g,\b}E^{\b,\a}
   -q^{(-1)^{[\b]}}E^{\b,\a}E^{\g,\b},~~1\leq\a<\b<\g\leq m+n.
   \label{non-simple2}
\end{eqnarray}
$\GL$ is a $Z_2$-graded triangular Hopf superalgebra. For the
Cartan and simple generators, the coproduct
$\Delta:\GL\rightarrow\GL\otimes \GL$ is defined by
\begin{eqnarray}
&&\Delta(E^{i,i})=1\otimes E^{i,i}+E^{i,i}\otimes
1,~i=1,\ldots,m+n,\label{copr-1}\\
&&\Delta(E^{j,j+1})=1\otimes E^{j,j+1}+E^{j,j+1}\otimes q^{h^j},\label{copr-2}\\
&&\Delta(E^{j+1,j})=q^{-h^{j}}\otimes E^{j+1,j}+E^{j+1,j}\otimes
1,\label{copr-3}
\end{eqnarray}
where $h^{j}=(-1)^{[j]}E^{j,j}-(-1)^{[j+1]}E^{j+1,j+1}$.
Throughout, we will use the following notation:
\begin{eqnarray}
E_{\a,\b}\equiv\Delta^{(N-1)}(E^{\a,\b})=({\rm id}\otimes
\Delta^{(N-2)})\Delta(E^{\a,\b}), \label{de:N-tensor}
\end{eqnarray}
for any generator $E^{\a,\b}$ $(\a,\b=1,\ldots,m+n)$ of $\GL$.

\subsection{Integrability of the model}

The $R$-matrix, $R\in \mbox{End}(V\otimes V)$, depends on the
difference of two spectral parameters $\l_1$ and $\l_2$ associated
with the two copies of $V$, and is, in this grading, given by
\cite{Per81,Kul82,Kul86}
\begin{eqnarray}
 &&R_{12}(\l_1,\l_2)=R_{12}(\l_1-\l_2)\nonumber\\
 &=&c_{12}\sum_{i=1}^m e_{i,i}\otimes e_{i,i}
  +\sum_{i=m+1}^{m+n} e_{i,i}\otimes e_{i,i}
  +a_{12}\sum_{i\ne j=1}^{m+n} e_{i,i}\otimes e_{j,j}\nonumber\\ &&
  \mbox{}
  +b^-_{12}\sum_{i>j=1}^{m+n}(-1)^{[j]}e_{i,j}\otimes e_{j,i}
  +b^+_{12}\sum_{j>i=1}^{m+n}(-1)^{[j]}e_{i,j}\otimes
  e_{j,i}, \label{R12} 
\end{eqnarray}
where
\begin{eqnarray}
&& a_{12}=a(\l_1,\l_2)\equiv {\sinh(\l_1-\l_2)\over
             \sinh(\l_1-\l_2+\eta)},\quad
   b_{12}^\pm=b^\pm(\l_1,\l_2)\equiv{e^{\pm(\l_1-\l_2)}\sinh\eta\over
          \sinh(\l_1-\l_2+\eta)},\quad\quad \nonumber\\
&& c_{12}=c(\l_1,\l_2)\equiv{\sinh(\l_1-\l_2-\eta)\over
\sinh(\l_1-\l_2+\eta)}, \label{de:abc}
\end{eqnarray}
and $\eta$ is the so-called crossing parameter. One can easily
check that the $R$-matrix satisfies the unitary relation
$R_{21}R_{12}=1.$ 

Let us introduce the $(N+1)$-fold tensor product space $V^{\otimes
N+1}$, whose components are labelled by $0,1\ldots,N$ from the
left to the right. As usual, the $0$-th space, denoted by $V_0$ (
$V_i$ for the $i$-th space), corresponds to the auxiliary space
and the other $N$ spaces constitute the quantum space $V^{\otimes
N}$. Moreover, for each factor space $V_i$, $i=0,\ldots,N$, we
associate a complex parameter $\x_i$. The parameter associated
with the $0$-th space is usually called the {\it spectral}
parameter which is set  to $\x_0=\l$ in this paper, and the other
parameters are called the {\it inhomogeneous} parameters. In this
paper we always assume that all the complex parameters $u$ and
$\{\x_i|i=1,\ldots,N\}$ be {\it generic} ones. Hereafter  we adopt
the standard notation: for any matrix $A\in {\rm End}(V)$,  $A_j$
(or $A_{(j)}$) is an embedding operator in the tensor product
space, which acts as $A$ on the $j$-th space and as an identity on
the other factor spaces; $R_{ij}=R_{ij}(\x_i,\x_j)$ is an
embedding operator of R-matrix in the tensor product space, which
acts as an identity on the factor spaces except for the $i$-th and
$j$-th ones.

The $R$-matrix satisfies the graded Yang-Baxter equation (GYBE)
\begin{equation}
R_{12}R_{13}R_{23}=R_{23}R_{13}R_{12}.
\end{equation}
In terms of the matrix elements defined by
\begin{equation}
R(\l)(v^{i'}\otimes
v^{j'})=\sum_{i,j}R(\l)^{i'j'}_{ij}(v^{i}\otimes v^{j}),
\end{equation}
the GYBE reads
\begin{eqnarray}
&&
\sum_{i',j',k'}R(\l_1-\l_2)^{i'j'}_{ij}R(\l_1-\l_3)^{i''k'}_{i'k}R(\l_2-\l_3)^{j''k''}_{j'k'}
    (-1)^{[j']([i']+[i''])}\nonumber\\
&=&\sum_{i',j',k'}R(\l_2-\l_3)^{j'k'}_{jk}R(\l_1-\l_3)^{i'k''}_{ik'}R(\l_1-\l_2)^{i''j''}_{i'j'}
    (-1)^{[j']([i]+[i'])}. \label{GYBE}
\end{eqnarray}

The quantum monodromy matrix $T(\l)$ of the supersymmetric chain
of length $N$ is defined as
\begin{eqnarray}
T(\l)=R_{0N}(\l,\x_N)R_{0N-1}(\l,\x_{N-1})_{\ldots}
     R_{01}(\l,\x_1), \label{de:T}
\end{eqnarray}
where the index 0 refers to the auxiliary space  and $\{\x_i\}$
are arbitrary inhomogeneous parameters depending on site $i$.

By the GYBE, one may prove that the monodromy matrix satisfies the
GYBE \bea
R_{00'}(\l-\m)T_0(\l)T_{0'}(\m)=T_{0'}(\m)T_{0}(\l)R_{00'}(\l-\m).
\label{eq:GYBE-1}\eea
 Define the transfer matrix
$t(\l)=str_0T(\l)$,
where $str_0$ denotes the supertrace over the auxiliary space.
Then the Hamiltonian of our model is given by
$H={d\ln t(\l)/ d\l}|_{\l=0}.$
This model is integrable thanks to the commutativity of the
transfer matrix for different parameters,
$[t(\l),t(\m)]=0,$
which can be verified by using the GYBE.

\subsection{$N$-site $R$-matrices}

Let $\s$ be an element of the permutation group ${\cal S}_{N+1}$.
We generalize the $R$-matrix (\ref{R12}) to the $N$-site
$R$-matrix with the help of $\s$ as follows. The concept of
$N$-site $R$-matrices was first introduced in
\cite{Maillet96,Albert00}.
\begin{Definition} One can define a mapping from ${\cal S}_{N+1}$ to ${\rm
End}(V_0\otimes{\cal{H}})$ which associate in a unique way an
element  $R^{\s}_{0\ldots N}\in {\rm End}(V_0\otimes {\cal{H}})$
to any element $\s$ of the permutation group ${\cal S}_{N+1}$. The
mapping has the following composition law
\begin{eqnarray}
 R^{\s\s'}_{0\ldots
N}={\cal P}^{\s}\,R^{\s'}_{0\ldots N}\,({\cal P}^{\s})^{-1}
\,R^{\s}_{0\ldots N}=R^{\s'}_{\s(0\ldots N)}\,R^{\s}_{0\ldots
N},~\forall \s,\s'\in \S_{N+1},\label{elemet-1}
\end{eqnarray}
where  ${\cal P}^{\s}$ is the $\Zb_2$-graded permutation operator
$${\cal P}^{\s}|i_0\rangle_{(0)}\ldots|i_N\rangle_{(N)}
=|i_0\rangle_{(\s(0))}\ldots|i_N\rangle_{(\s(N))}.$$

For any elementary permutation $\s_{j}$ with
$\s_j(j,j+1)=(j+1,j)$,
$j=0,\ldots,N$,  $R^{\s_j}_{0\ldots N}=R_{j\,j+1}.$
\end{Definition}
 \vskip0.1in

From the definition, one may prove the following properties of the
map $R^\s_{0\ldots N}$:

\begin{itemize}
\item  Uniqueness. For any element $\s\in {\cal S}_{N+1}$, the
corresponding $R^{\s}_{0\ldots N}$ can be constructed through
(\ref{elemet-1}) as follows. Let $\s$ be decomposed in a minimal
way in terms of elementary permutation as
$\s=\s_{\b_1}\ldots\s_{\b_p}$ where the positive integer $p$ is
the length of $\s$. The composition law enables one to obtain  the
expression of the associated $R^{\s}_{0\ldots N}$. The GYBE and
the  unitary relation guarantee the uniqueness of $R^{\s}_{0\ldots
N}$.

\item  $R^{id}_{0\ldots N}=id$. By using the unitary relation
$R_{ij}R_{ji}=1$, this property can be easily proved.

\item  For the cyclic permutation $\s_c'=\s_0\s_1\ldots\s_N$ of
the group ${\cal S}_{N+1}$, the $N$-site $R$-matrix
$R^{\s_c'}_{0\ldots N}$ gives the
 monodromy matrix $T(\l)$
of the $U_q(gl(m|n))$ spin chain on an $N$-site lattice:
 \begin{equation}
R^{\s_c'}_{0\ldots
N}=R_{0\,N}R_{01\,N-1}\ldots\,R_{01}=T(\l)\equiv
T_0(\l)=T_{0,1\ldots N}(\l) .\label{Mon}
\end{equation}
%

\item  For any $\s\in{\cal S}_N$ which acts on the quantum space
${\cal H}$, by using the GYBE, one may prove
\begin{eqnarray}
 R_{1\ldots N}^{\sigma}T_{0,1\ldots N}
  =T_{0,\sigma'(1\ldots N)}R_{1\ldots N}^{\sigma}.
     \label{eq:RT-sigma}
\end{eqnarray}
Moreover,  let $\s_c=\s_1\ldots\s_{N-1}$ be the cyclic
permutation. We have
\begin{eqnarray}
R^{{\s_c}^k}_{1\ldots N}T_{0,1\ldots N}(\l)
 =T_{0,k+1\ldots N1\ldots k}(\l)R^{{\s_c}^k}_{1\ldots N}.
 \label{eq:RT}
\end{eqnarray}

\end{itemize}


\sect{Some properties of the monodromy matrix elements}
\subsection{Recursive relations}

The monodromy matrix $T(\l)$ (\ref{de:T}) may be decomposed in
terms of the basis of ${\rm End}(V_0)$ as
 \begin{eqnarray}
T(\l)=\sum_{i,j=1}^{m+n}T_{i,j}(\l)E^{i,j}_{(0)}\equiv
\sum_{i,j=1}^{m+n} T_{i,j}(\l)e_{i,j}, \label{def-T}
\end{eqnarray}
where the matrix elements $T_{i,j}(\l)$ are operators acting on
the quantum space ${\cal{H}}$ and have  the $\Zb_2$-grading:
$[T_{i,j}(\l)]=[e_{i,j}]=[i]+[j]$.

For the entries $T_{m+n,m+n-l}(\l)$ of the monodromy matrix, we
have the following theorem \cite{zsy0503}:

\vskip0.1in
\begin{Theorem} \label{Theo-1}
The matrix elements $T_{m+n,m+n-l}(\l)$ and $T_{m+n-l,m+n}(\l)$
$(l=1,\ldots,m+n-1)$ of the monodromy matrix can be expressed  in
terms of $T_{m+n,m+n}(\l)$ and the generators of $\GL$ by the
following recursive relations: \bea
 &&T_{m+n,m+n-l}(\l)=\lt(q^{-(-1)^{[m+n]}}
 E_{m+n-l,m+n}T_{m+n,m+n}(\l)\rt.
 \no\\ &&\quad\mbox{}\lt.
 -T_{m+n,m+n}(\l)E_{m+n-l,m+n}\rt)
q^{-\sum_{k=1}^{l}H_{m+n-k}}\no\\
&&-\sum_{\a=1}^{l-1}(1-q^{-2(-1)^{[m+n-\a]}})T_{m+n,m+n-\a}(\l)E_{m+n-l,m+n-\a}
q^{-\sum_{k=\a+1}^{l}H_{m+n-k}}.\no\\ \label{Recursive-1}\eea
 \begin{eqnarray}
&&T_{m+n-l,m+n}(\l)=(-1)^{[m+n]+[m+n-l]}q^{\sum_{k=1}^lH_{m+n-k}}\nonumber\\
&&\quad \mbox{}\times
 \lt(q^{(-1)^{[m+n]}}T_{m+n,m+n}(\l)E_{m+n,m+n-l}
  -E_{m+n,m+n-l}T_{m+n,m+n}(\l)\rt)
  \nonumber\\ &&\quad\mbox{}-
  \sum_{\a=1}^{l-1}(-1)^{([m+n]+[m+n-l])([m+n-l]+[m+n-l+\a])}
  (1-q^{2(-1)^{[m+n-l+\a]}})\no\\
&&\quad\quad\times
  q^{\sum_{k=l-\a+1}^{l}H_{m+n-k}}
  E_{m+n-l+\a,m+n-l}T_{m+n-l+\a,m+n}(\l),\label{Recursive-2}
\end{eqnarray}
where $E_{i,j}$ is the $N$-site $\GL$ generator which is given by
local generator $E^{i,j}_{(k)}$ with the help of
(\ref{de:N-tensor}), and
$H_j=(-1)^{[j]}E_{j,j}-(-1)^{[j+1]}E_{j+1,j+1}~(j=1,\ldots,m+n-1)$.

\end{Theorem}

\noindent We call the second terms in the  R.H.S. of
(\ref{Recursive-1}) and (\ref{Recursive-2})  {\it quantum
correction term}, which vanishes in the rational limit
($q\rightarrow 1$). Moreover, such a nontrivial correction term
only occurs in the higher rank models (i.e., when $m+n\geq 3$). In
the rational limit: $q\rightarrow 1$, (\ref{Recursive-1}) and
(\ref{Recursive-2}) reduces to the (anti)commutation relations
used in \cite{Albert00,zsy0502}. The detailed proof for this
theorem may be found in \cite{zsy0503}. Here we give two examples
to
illustrate the theorem. \\
\begin{itemize}
\item For $m=2,n=0$, i.e. the $U_q(gl(2|0))$ model:
\begin{eqnarray}
 && T_{2,1}=[qE_{1,2}T_{2,2}-T_{2,2}E_{1,2}]q^{H_1},\\
 && T_{1,2}=q^{H_1}[q^{-1}T_{2,2}E_{2,1}-E_{2,1}T_{2,2}].
\end{eqnarray}
\item For $m=2,n=1$, i.e. the $q$-deformed supersymmetric $t-J$
model:
 \bea
&&T_{3,2}(\l)=\lt[q^{-1}E_{2,3}T_{3,3}(\l)-T_{3,3}(\l)E_{2,3}\rt]q^{-H_2},\\
&&T_{2,3}{\l}=-q^{H_2}\lt[q~T_{3,3}(\l)E_{3,2}-E_{3,2}T_{3,3}(\l)\rt],\\
&&T_{3,1}(\l)=\lt[q^{-1}E_{1,3}T_{3,3}(\l)-T_{3,3}(\l)E_{1,3}\rt]q^{-H_2-H_1}
 \no\\&&\quad\quad\quad\quad\quad
-(1-q^2)T_{3,2}(\l)E_{1,2}q^{-H_1},\\
&&T_{1,3}(\l)=-q^{H_1+H_2}\lt[q~T_{3,3}(\l)E_{3,1}
  -E_{3,1}T_{3,3}(\l)\rt]
   \no\\ &&\quad\quad\quad\quad\quad
   -(1-q^{-2})q^{H_1}E_{2,1}T_{2,3}(\l).\eea

\end{itemize}

\subsection{Representation of the local operators}

In \cite{Kitanine98}, Kitanine et al constructed the local spin
operators of the inhomogeneous spin-1/2 XXX and XXZ Heisenberg
chains in terms of the corresponding monodromy matrix elements.
The results were generalized to more general cases in
\cite{Korepin99,Maillet99}.
%
\begin{Theorem}
For the monodromy matrix constructed by the $\GL$ $R$-matrix which
is a solution of the GYBE and satisfies the properties: i.)
Regularity. $R_{ij}(\l_i,\l_j=\l_i)={\cal P}_{ij}$ and ii.)
 $R(\l,\m)^{i'j'}_{i\,j}
  =(-1)^{[i]+[j]+[i']+[j']}R(\l,\m)^{i'j'}_{i\,j}$, the local generator
$E^{\a,\b}_{(\k)}$, which acts on the given site $\k$, can be
represented  in terms of the entries of the monodromy matrix as
the following formula,
\begin{eqnarray}
 E_{(\k)}^{\a,\b}=(-1)^{[\a][\b]}
  \prod_{j=1}^{\k-1}str_0(T_{0,1\ldots N}(\x_j))\,
  T_{\a,\b}(\x_\k)\,
  \prod_{j=\k+1}^{N}str_0(T_{0,1\ldots N}(\x_j)). \label{eq:E-T}
\end{eqnarray}
\end{Theorem}
Here  ${\cal P}$ is the superpermutation operator, i.e. ${\cal
P}(|x\rangle\otimes|y\rangle)
 =(-1)^{[x][y]}(|y\rangle\otimes|x\rangle)$.
In terms of the generators $E^{\a,\b}_{(k)}$, ${\cal P}_{ij}$ can
be written by
$ {\cal
P}_{ij}=\sum_{\a,\b}(-1)^{[\a]}E^{\a,\b}_{(i)}E^{\b,\a}_{(j)}$.

 \vskip0.1in

To prove this theorem, one considers the supertrace
$str_0(X_0T_{0,\k\ldots N1\ldots \k-1}(\x_\k))$, where $X_0\in
\GL$ and $X_0=\sum_{\a,\b=1}^{m+n}X_{\a,\b}E_{(0)}^{\b,\a}$,
\begin{eqnarray}
 && str_0(X_0T_{0,\k\ldots N1\ldots\k-1}(\x_\k))
 \nonumber\\&&
 =str_0(X_0R_{0\k-1}(\x_n,\x_{\k-1})\ldots
  R_{01}(\x_n,\x_{1}))R_{0N}(\x_n,\x_{N})\ldots
  R_{0\k+1}(\x_n,\x_{\k+1}){\cal P}_{0\k})
 \nonumber\\&&
 =str_0(X_0{\cal P}_{0\k})R_{\k\k-1}\ldots
  R_{\k1}R_{\k N}\ldots R_{\k\k+1}
 \nonumber\\&&
 =\sum_{\a,\b=1}^{m+n}(-1)^{[\a]+[\b]}X_{\a,\b} E_{(\k)}^{\b,\a}
  ~R^{\s_c}_{\s_c^{\k-1}(1\ldots N)}. \label{eq:E-T-1}
\end{eqnarray}
where $\s_c=\s_1\ldots\s_{N-1}\,\, (\s\in {\cal S}_N)$ and in the
derivation of (\ref{eq:E-T-1}), we have used the decomposition law
(\ref{elemet-1}).

Similarly, one may prove the following useful relations:
\begin{eqnarray}
 && str_0(T_{0,1\ldots N}(\x_\k))=R^{\s_c}_{\s_c^{\k-1}(1\ldots N)},
 \quad\quad
  \prod_{j=1}^{\k}str_0(T_{0,1\ldots N}(\x_j))
 =R^{\s_c^\k}_{1\ldots N},\no\\
 &&\prod_{j=\k+1}^{N}str_0(T_{0,1\ldots N}(\x_j))
 =\left(R^{\s_c^\k}_{1\ldots N}\right)^{-1}.
 \label{eq:E-T-4}
\end{eqnarray}

On the other hand, with the help of the decomposition laws
(\ref{elemet-1}), (\ref{eq:RT}) and (\ref{eq:E-T-4}), we have
\begin{eqnarray}
 && str_0(X_0T_{0,\k\ldots N1\ldots\k-1}(\x_\k))
 \nonumber\\&&
 =str_0\left(X_0T_{0,\k\ldots N1\ldots\k-1}(\x_\k)
 R^{\s_c^{\k-1}}_{1\ldots N}
 \left(R^{\s_c^{\k-1}}_{1\ldots N}\right)^{-1}\right)
 \nonumber\\&&
 =R^{\s_c^{\k-1}}_{1\ldots N}\cdot
  str_0\left(X_0T_{0,1\ldots N}(\x_\k)\right) \cdot
 \left(R^{\s_c^{\k-1}}_{1\ldots N}\right)^{-1}
 \nonumber\\&&
 =(-1)^{[\a]}
  \prod_{j=1}^{\k}str_0(T_{0,1\ldots N}(\x_j))\cdot
  T_{\a,\b}\cdot
  \prod_{j=\k}^{N}str_0(T_{0,1\ldots N}(\x_j)).
  \label{eq:E-T-5}
\end{eqnarray}
Then comparing (\ref{eq:E-T-1}) with  (\ref{eq:E-T-5}), and
considering (\ref{eq:E-T-4}) and the supertranspose property
$A_{ij}=(-1)^{[j]([i]+[j])}A_{ji}$, one arrives at (\ref{eq:E-T}),


\section{Factorizing  F-matrices and their inverses}
\label{F-matrix} \setcounter{equation}{0}


In \cite{Maillet96}, Maillet {\it et al} found that the
$R$-matrices for the XXX and XXZ Heisenberg spin chain models are
factorized in terms of the $F$-matrices. The results were
generalized to $gl(n)$ spin chain system by Albert et al
\cite{Albert00}, where the authors constructed the factorizing
$F$-matrices (Drinfeld twists) explicitly on the $N$-fold tensor
product space.

Let $\S_{N}$ be the permutation group associated with the indices
$(1,\ldots,N)$  and $R^{\s}_{1\ldots N}$ the $N$-site $R$-matrix
associated with $\s\in\S_N$. $R^{\s}_{1\ldots N}$ acts
non-trivially on the quantum space ${\cal{H}}$ and trivially  (i.e
as an identity) on the auxiliary space.

\vskip0.in

\begin{Definition} The $N$-site F-matrix $F_{1\ldots N}(\x_1,\ldots,\x_N)$
is an operator in ${\rm End}({\cal{H}})$ and satisfies the
following three properties:  I.) lower-triangularity; II.)
non-degeneracy;  III.) factorization, namely,\bea
F_{\s(1)\ldots\s(N)}(\x_{\s(1)},\ldots,\x_{\s(N)})\,
R^{\s}_{1\ldots N}=F_{1\ldots N}(\x_1,\ldots,\x_N),~\forall \s\in
\S_N.\eea
\end{Definition}
\vskip0.1in


\begin{Proposition}
The $N$-site factorizing $F$-matrix for the  $\GL$ supersymmetric
model, given by
\begin{eqnarray}
F_{1\ldots N}\equiv F_{1\ldots
N}(\x_1,\ldots,\x_N)=\sum_{\sigma\in {\cal S}_N}
   {\sum_{\alpha_{\sigma(1)}\ldots\alpha_{\sigma(N)}=1}^{m+n}}^{\hspace{-0.6truecm}*}
   \hspace{0.6truecm}\prod_{j=1}^N P_{\sigma(j)}^{\alpha_{\sigma(j)}}
   S(\sigma,\alpha_\sigma)R_{1\ldots N}^\sigma, \label{de:F}
\end{eqnarray}
satisfies the properties I, II and III in the definition 2.
\end{Proposition}
Here, $P^{\a}_i$ is the embedding of the project operator $P^{\a}$
in the $i$-th space with $(P^{\a})_{kl}=\d_{kl}\d_{k\a}$,
 $S(\sigma,\alpha_\sigma)$ is a c-number function of
$\s,\a_\s$ and the element $c_{ij}$ of the R-matrix
\begin{eqnarray}
S(\sigma,\alpha_\sigma)\equiv
\exp\lt\{\frac{1}{2}\sum_{l>k=1}^N\lt(1-(-1)^{[\a_{\s(k)}]}\rt)\,
\delta_{\alpha_{\sigma(k)},\alpha_{\sigma(l)}}
    \ln(1+c_{\sigma(k)\sigma(l)})\rt\}\label{F-1}
\end{eqnarray}
and the sum $\sum^*$ is defined by
\begin{eqnarray}
&& \alpha_{\sigma(i+1)}\geq \alpha_{\sigma(i)}\,
              (\sigma(i+1)>\sigma(i));
 \quad \alpha_{\sigma(i+1)}> \alpha_{\sigma(i)}\,
              (\sigma(i+1)<\sigma(i)). \label{cond:F}
\end{eqnarray}

Here we outline the proof given in \cite{zsy04,zsy0503}. The
definition of $F_{1\ldots N}$ (\ref{de:F}) and the summation
condition (\ref{cond:F}) imply that $F_{1\ldots N}$ is a
lower-triangular matrix. Moreover, one can easily check that the
$F$-matrix is non-degenerate because all diagonal elements are
non-zero.

We now prove that the $F$-matrix (\ref{de:F}) satisfies the
property III. Any given permutation $\sigma\in {\cal S}_N$ can be
decomposed into elementary ones of the group ${\cal S}_N$ as
$\sigma=\sigma_{i_1}\ldots \sigma_{i_k}$.  By (\ref{elemet-1}), we
have, if the property III holds for any elementary permutation
$\sigma_i$,
\begin{eqnarray}
&&F_{\sigma(1\ldots N)}R^{\sigma}_{1\ldots N}
 = F_{\sigma_{i_1}\ldots\sigma_{i_k}(1\ldots N)}
     R^{\sigma_{i_k}}_{\sigma_{i_1}\ldots\sigma_{i_{k-1}}(1\ldots N)}
     R^{\sigma_{i_{k-1}}}_{\sigma_{i_1}\ldots\sigma_{i_{k-2}}(1\ldots N)}
     \ldots
     R^{\sigma_{i_1}}_{1\ldots N}\nonumber\\
 &=&F_{\sigma_{i_1}\ldots\sigma_{i_{k-1}}(1\ldots N)}
     R^{\sigma_{i_{k-1}}}_{\sigma_{i_1}\ldots\sigma_{i_{k-2}}(1\ldots N)}
     \ldots
     R^{\sigma_{i_1}}_{1\ldots N}
  =\ldots
     =F_{\sigma_{i_1}(1\ldots N)}R^{\sigma_{i_1}}_{1\ldots N}=F_{1\ldots N}.
     \no
\end{eqnarray}

For the elementary permutation $\sigma_i$, we have
\begin{eqnarray}
 &&F_{\sigma_i(1\ldots N)}R^{\sigma_i}_{1\ldots N}
 =\sum_{\sigma\in {\cal S}_N}
   \sum_{\alpha_{\sigma_i\sigma(1)}\ldots\alpha_{\sigma_i\sigma(N)}}^{\quad\quad *}
   \prod_{j=1}^N P_{\sigma_i\sigma(j)}^{\alpha_{\sigma_i\sigma(j)}}
  S(\sigma_i\sigma,\alpha_{\sigma_i\sigma})R_{\sigma_i(1\ldots N)}^\sigma
   R_{1\ldots N}^{\sigma_i} \nonumber\\
 &=&\sum_{\sigma\in {\cal S}_N}
   \sum_{\alpha_{\sigma_i\sigma(1)}\ldots\alpha_{\sigma_i\sigma(N)}}^{\quad\quad *}
   \prod_{j=1}^N P_{\sigma_i\sigma(j)}^{\alpha_{\sigma_i\sigma(j)}}
   S(\sigma_i\sigma,\alpha_{\sigma_i\sigma})
   R_{1\ldots N}^{\sigma_i\sigma} \nonumber\\
 &=&\sum_{\tilde\sigma\in {\cal S}_N}
   \sum_{\alpha_{\tilde\sigma(1)}\ldots\alpha_{\tilde\sigma(N)}}^{\quad\quad *(i)}
   \prod_{j=1}^N P_{\tilde\sigma(j)}^{\alpha_{\tilde\sigma(j)}}
   S(\tilde\sigma,\alpha_{\tilde\sigma})R_{1\ldots N}^{\tilde\sigma}, \label{eq:FR-F}\nonumber\\
\end{eqnarray}
where $\tilde\sigma=\sigma_i\sigma$, and the summation sequences
of $\alpha_{\tilde\sigma}$ in ${\sum^*}^{(i)}$ now has the form
\begin{eqnarray}
 \alpha_{\tilde\sigma(j+1)}\geq \alpha_{\tilde\sigma(j)}\,
              (\sigma_i\tilde\sigma(j+1)>\sigma_i\tilde\sigma(j));
\quad \alpha_{\tilde\sigma(j+1)}> \alpha_{\tilde\sigma(j)}\,
              (\sigma_i\tilde\sigma(j+1)<\sigma_i\tilde\sigma(j)). \nonumber\\
              \label{cond:FR-F}
\end{eqnarray}
Comparing (\ref{cond:FR-F}) with (\ref{cond:F}), we find that the
only difference between them is the transposition $\sigma_i$
factor in the ``if" conditions.  For a given $\tilde\sigma\in
{\cal S}_N$ with $\tilde\sigma(j)=i$ and $\tilde\sigma(k)=i+1$, we
now examine how the elementary transposition $\sigma_i$ will
affect the inequalities (\ref{cond:FR-F}). If $|j-k|>1$, then
$\sigma_i$ does not affect the sequence of $\alpha_{\tilde\sigma}$
at all, that is, the sign of inequality $``>"$ or ``$\geq$"
between two neighboring root indexes is unchanged with the action
of $\sigma_i$. If $|j-k|=1$, then in the summation sequences of
$\alpha_{\tilde\sigma}$, when $\tilde\sigma(j+1)=i+1$ and
$\tilde\sigma(j)=i$, sign ``$\geq$" changes to $``>"$, while when
$\tilde\sigma(j+1)=i$ and $\tilde\sigma(j)=i+1$, $``>"$ changes to
``$\geq$". Thus (\ref{cond:F}) and (\ref{eq:FR-F}) differ only
when equal labels $\alpha_{\tilde\sigma}$ appear. With the help of
the relation $c_{21}c_{12}=1$, we can prove that in this case the
product $F_{\sigma_i(1\ldots N)}R^{\sigma_i}_{1\ldots N}$ still
equals to $F_{1\ldots N}$ (one sees in \cite{zsy04} for detailed
proof). Thus, we obtain
\begin{eqnarray}
R_{1\ldots N}^\sigma(\x_1,\ldots,\x_N)
 =F^{-1}_{\sigma(1\ldots N)}(\x_{\sigma(1)},\ldots,
   \x_{\sigma(N)})F_{1\ldots N}(\x_1,\ldots,\x_N),
\end{eqnarray}
Therefore the factorizing $F$-matrix $F_{1\ldots N}$ of
$U_q(gl(m|n))$ is proved to satisfy all three properties.

From the expression of the $F$-matrix, one knows that it has an
even grading, i.e., $ [F_{1\ldots N}]=0.$


The non-degenerate property of the $F$-matrix implies that we can
find the inverse matrix $F^{-1}_{1\ldots N}$.

\vskip0.1in
\begin{Proposition}
The inverse of the $F$-matrix is given by
\begin{equation}
F^{-1}_{1\ldots N}=F^*_{1\ldots
N}\prod_{i<j}\Delta_{ij}^{-1},\label{Prop-4}
\end{equation}
where
\begin{eqnarray}
F^*_{1\ldots N}&=&\sum_{\sigma\in {\cal S}_N}
   {\sum_{\alpha_{\sigma(1)}\ldots\alpha_{\sigma(N)}=1}^{m+n}}
   ^{\hspace{-0.6truecm}\vspace{-.4truecm} **}
   S(\sigma,\alpha_\sigma)R_{\sigma(1\ldots N)}^{\sigma^{-1}}
   \prod_{j=1}^N P_{\sigma(j)}^{\alpha_{\sigma(j)}},
    \label{de:F*}
\end{eqnarray}
and
\begin{eqnarray}
[\Delta_{ij}]^{\beta_i\beta_j}_{\alpha_i\alpha_j} =
 \delta_{\alpha_i\beta_i}\delta_{\alpha_j\beta_j}\left\{
 \begin{array}{cl}
  a_{ij}&\, \,
   \mbox{if} \ \alpha_i>\alpha_j\\
  a_{ji}& \, \,
   \mbox{if} \ \alpha_i<\alpha_j,\\
  1& \, \, \mbox{if}\ \alpha_i=\alpha_j=m+1,\ldots,m+n, \\ \displaystyle
  (1+c_{ij})(1+c_{ji})& \, \,
  \mbox{if}\ \alpha_i=\alpha_j=1,\ldots,m.
  \end{array}\right. \label{de:F-Inv}
\end{eqnarray}

\end{Proposition}
Here the sum $\sum^{**}$ is taken over all possible $\alpha_i$
which satisfies the following non-increasing constraints:
\begin{eqnarray}
 \alpha_{\sigma(i+1)}\leq \alpha_{\sigma(i)}\,
              (\sigma(i+1)<\sigma(i)); 
\quad \alpha_{\sigma(i+1)}< \alpha_{\sigma(i)}\,
              \lt(\sigma(i+1)>\sigma(i)\rt). \label{cond:F*}
\end{eqnarray}

\vskip0.1in

We outline the proof in \cite{zsy04,zsy0503}. We compute the
product of $F_{1\ldots N}$ and $F^*_{1\ldots N}$. Substituting
(\ref{de:F}) and (\ref{de:F*}) into the product, we have
\begin{eqnarray}
 &&F_{1\ldots N}F^*_{1\ldots N}
 =\sum_{\sigma\in {\cal S}_N}\sum_{\sigma'\in {\cal S}_N}
    \sum^{\quad\quad *}_{\alpha_{\sigma_1}\ldots\alpha_{\sigma_N}}
    \sum^{\quad\quad **}_{\beta_{\sigma'_1}\ldots\beta_{\sigma'_{N}}}
    S(\sigma,\alpha_\sigma)S(\sigma',\beta_{\sigma'})
    \nonumber\\ &&\times
    \prod_{i=1}^N P_{\sigma(i)}^{\alpha_{\sigma(i)}}
    R^{\sigma}_{1\ldots N}R^{{\sigma'}^{-1}}_{\sigma'(1\ldots N)}
    \prod_{i=1}^N P_{\sigma'(i)}^{\beta_{\sigma'(i)}} \nonumber\\
 &=&\sum_{\sigma\in {\cal S}_N}\sum_{\sigma'\in {\cal S}_N}
    \sum^{\quad\quad *}_{\alpha_{\sigma_1}\ldots\alpha_{\sigma_N}}
    \sum^{\quad\quad **}_{\beta_{\sigma'_1}\ldots\beta_{\sigma'_{N}}}
    S(\sigma,\alpha_\sigma)S(\sigma',\beta_{\sigma'})
    \prod_{i=1}^N P_{\sigma(i)}^{\alpha_{\sigma(i)}}
    R^{{\sigma'}^{-1}\sigma}_{\sigma'(1\ldots N)}
    \prod_{i=1}^N P_{\sigma'(i)}^{\beta_{\sigma'(i)}}. \no
\end{eqnarray}
To evaluate the R.H.S., we examine the matrix element of the
$R$-matrix
$\left(R^{{\sigma'}^{-1}\sigma}_{\sigma'(1\ldots N)}\right)
  ^{\alpha_{\sigma(N)}\ldots\alpha_{\sigma(1)}}
  _{\beta_{\sigma'(N)}\ldots\beta_{\sigma'(1)}}.$
Note that the sequence $\{ \alpha_{\sigma}\}$ is non-decreasing
and $\{\beta_{\sigma'}\}$ is non-increasing. Thus the
non-vanishing condition of this $R$-matrix element requires that
$\alpha_{\sigma}$ and $\beta_{\sigma'}$ satisfy the relations
$\beta_{\sigma'(N)}=\alpha_{\sigma(1)},\ldots,
\beta_{\sigma'(1)}=\alpha_{\sigma(N)}.$
Then by using the sum conditions (\ref{cond:F}), (\ref{cond:F*})
and the existence condition of the elements of the elementary
$R$-matrix $R(\l,\m)^{i'j'}_{ij}$, i.e. $i+j=i'+j'$, one can
verify \cite{Albert00} that the relations between the roots $\b$
and $\a$ are fulfilled only if
$\sigma'(N)=\sigma(1),\ldots,\sigma'(1)=\sigma(N).$
Let $\bar\sigma$ be the maximal element of the ${\cal S}_N$ which
reverses the site labels
$\bar\sigma(1,\ldots,N)=(N,\ldots,1).$
Then we have
$\sigma'=\sigma\bar\sigma.$ 
Therefore we have
\begin{eqnarray}
 F_{1\ldots N}F^*_{1\ldots N}
 &=&\sum_{\sigma\in {\cal S}_N}
    \sum^{\quad\quad *}_{\alpha_{\sigma_1}\ldots\alpha_{\sigma_N}}
    S(\sigma,\alpha_\sigma)S(\sigma,\alpha_\sigma)
    \prod_{i=1}^N P_{\sigma(i)}^{\alpha_{\sigma(i)}}
    R^{\bar\sigma}_{\sigma(N\ldots 1)}
    \prod_{i=1}^N P_{\sigma(i)}^{\alpha_{\sigma(i)}}\label{eq:FF*-2}.
\end{eqnarray}
The decomposition of $R^{\bar\sigma}$ in terms of elementary
$R$-matrices is unique module the GYBE. One reduces from
(\ref{eq:FF*-2}) that $FF^*$ is a diagonal matrix:
$F_{1\ldots N}F^*_{1\ldots N}=\prod_{i<j}\Delta_{ij}.$
Then (\ref{Prop-4}) is a simple consequence of the above equation.


\section{Monodromy matrix in the $F$-basis}
\label{F-B} \setcounter{equation}{0}

In the previous section, we have given  the $F$-matrix and its
inverse which act on the quantum space ${\cal{H}}$. The
non-degeneracy of the $F$-matrix means that its column vectors
also form a complete basis of ${\cal{H}}$, which is called the
$F$-basis. In this section, we study the generators of
$U_q(gl(m|n))$ and the elements of the monodromy matrix in the
$F$-basis.


Introduce the generators in the $F$-basis: $
\tilde{E}_{i,j}=F_{1\ldots N}E_{i,j}F^{-1}_{1\ldots
N},~(i,j=1,\ldots,m+n).$ Then,

\vskip0.1in
\begin{Theorem}  In the $F$-basis the Cartan and the simple generators of
$U_q(gl(m|n))$ are given by\bea
\tilde{E}_{i,i}&=&E_{i,i}=\sum_{k=1}^{N}E^{i,i}_{(k)},~i=1,\ldots,m+n,
\label{H-form-1}\\
\tilde{E}_{j,j+1}&=&\sum_{k=1}^{N}E^{j,j+1}_{(k)}\otimes_{\g\neq
k}G^{j,j+1}_{(\g)}(k,\g),~j=1,\ldots,m+n-1,\\
\tilde{E}_{j+1,j}&=&\sum_{k=1}^{N}E^{j+1,j}_{(k)}\otimes_{\g\neq
k}G^{j+1,j}_{(\g)}(k,\g), ~j=1,\ldots,m+n-1.\eea Here the diagonal
matrices $G^{\g,\g\pm 1}_{(j)}(i,j)$ are:\begin{itemize} \item For
$1<\g+1\leq m$, \bea
(G^{\g,\g+1}_{(j)}(i,j))_{kl}&=&\d_{kl}\lt\{\begin{array}{ll}
2e^{-\eta}\cosh\eta,&k=\g,\\(2a_{ij}\cosh\eta)^{-1}\,e^{\eta},
&k=\g+1,\\
1,&{\rm otherwise},\end{array}\rt.\\[4pt]
(G^{\g+1,\g}_{(j)}(i,j))_{kl}&=&\d_{kl}\lt\{\begin{array}{ll}
2e^{-\eta}\cosh\eta,&k=\g+1,\\(2a_{ji}\cosh\eta)^{-1}\,e^{\eta},
&k=\g,\\
1,&{\rm otherwise},\end{array}\rt. \eea

\item For $\g= m$, \bea
(G^{\g,\g+1}_{(j)}(i,j))_{kl}&=&\d_{kl}\lt\{\begin{array}{ll}
2e^{-\eta}\cosh\eta,&k=\g,\\e^{-\eta},
&k=\g+1,\\
1,&{\rm otherwise},\end{array}\rt.\\[4pt]
(G^{\g+1,\g}_{(j)}(i,j))_{kl}&=&\d_{kl}\lt\{\begin{array}{ll}
(2a_{ji}\cosh\eta)^{-1}e^{\eta},&k=\g,\\(a_{ji})^{-1}\,e^{\eta},
&k=\g+1,\\
1,&{\rm otherwise},\end{array}\rt. \eea

\item For $1+m\leq\g< m+n$, \bea
(G^{\g,\g+1}_{(j)}(i,j))_{kl}&=&\d_{kl}\lt\{\begin{array}{ll}
(a_{ij})^{-1}\,e^{\eta},&k=\g,\\e^{-\eta},
&k=\g+1,\\
1,&{\rm otherwise},\end{array}\rt.\\[4pt]
(G^{\g+1,\g}_{(j)}(i,j))_{kl}&=&\d_{kl}\lt\{\begin{array}{ll}
(a_{ji})^{-1}\,e^{-\eta},&k=\g+1,\\e^{\eta},
&k=\g,\\
1,&{\rm otherwise}.\end{array}\rt. \eea
\end{itemize}
\end{Theorem}

The proof for this theorem can be found in \cite{zsy04,zsy0503}.
This theorem plays an important role in the construction of the
symmetric representations of the creation (annihilation)
operators.


Among the matrix elements of the monodromy matrix $T_{i,j}(\l)$,
the operators $T_{m+n,m+n-l}(\l)$ and $T_{m+n-l,m+n}(\l)$
($l=1,\ldots,m+n-1$) are called creation and annihilation
operators, respectively, and  are usually denoted by
\begin{eqnarray}
&&C_{m+n-l}(\l)=T_{m+n,m+n-l}(\l),
~~B_{m+n-l}(\l)=T_{m+n-l,m+n}(\l).
 \end{eqnarray}
In the $F$-basis, they become
\begin{eqnarray}
\tilde{C}_{m+n-l}(\l)
 =F_{1\ldots N}C_{m+n-l}(\l)F^{-1}_{1\ldots N}, ~~~~
\tilde B_{m+n-l}(\l)
 =F_{1\ldots N}B_{m+n-l}(\l)F^{-1}_{1\ldots N}.\no\\ \label{Creation}
\end{eqnarray}
 Let us denote
$T_{m+n,m+n}(\l)$ by $D(\l)$ and the corresponding operator in the
$F$-basis by $ \tilde{D}(\l)=F_{1\ldots N}D(\l)F^{-1}_{1\ldots
N}. $

\vskip0.1in

\begin{Proposition} $\tilde{D}(\l)$ is a diagonal
matrix given by
\begin{eqnarray}
\tilde D(\l)=\otimes_{i=1}^N
  \mbox{diag}\left(a_{0i},\ldots,a_{0i},1\right)_{(i)}.
  \label{eq:T33-tilde}
\end{eqnarray}
\end{Proposition}

\vskip0.1in

This can be proved as follows \cite{zsy0503}. From (\ref{def-T}),
we derive that \bea
D(\l)P^{m+n}_{0}=T_{m+n,m+n}(\l)e_{m+n,m+n}=P^{m+n}_{0}T_{0,1\ldots
N}(\l)P^{m+n}_{0}.\eea Acting the $F$-matrix  from the left on the
both sides of the above equation, we have
\begin{eqnarray}
 F_{1\ldots N}D(\l)P^{m+n}_0
 &=&\sum_{\sigma\in {\cal S}_N}
    \sum_{\alpha_{\sigma(1)}\ldots\alpha_{\sigma(N)}}^{\quad\quad*}
    S(\sigma,\alpha_\sigma)\prod_{i=1}^N
    P_{\sigma(i)}^{\alpha_{\sigma}}R^\sigma_{1\ldots N}
    P_0^{m+n} T_{0,1\ldots N}(\l)P_0^{m+n} \nonumber\\
 &=&\sum_{\sigma\in {\cal S}_N}
    \sum_{\alpha_{\sigma(1)}\ldots\alpha_{\sigma(N)}}^{\quad\quad*}
    S(\sigma,\alpha_\sigma)\prod_{i=1}^N
    P_{\sigma(i)}^{\alpha_{\sigma}}
    P_0^{m+n} T_{0,\sigma(1\ldots N)}(\l)P_0^{m+n}
    R^\sigma_{1\ldots N}.\no\\
\end{eqnarray}
Following \cite{Albert00}, we can split the sum $\sum^*$ according
to the number of occurrences of the index $m+n$.
\begin{eqnarray}
F_{1\ldots N}D(\l)P^{m+n}_0
 &=&\sum_{\sigma\in {\cal S}_N}
    \sum_{k=0}^N
    \sum_{\alpha_{\sigma(1)}\ldots\alpha_{\sigma(N)}}^{\quad\quad*}
    S(\sigma,\alpha_\sigma)
    \prod_{j=N-k+1}^N \delta_{\alpha_{\sigma(j)},{m+n}}
    P_{\sigma(j)}^{\alpha_{\sigma(j)}} \nonumber\\ &&\times
    \prod_{j=1}^{N-k}P_{\sigma(j)}^{\alpha_{\sigma(j)}}
    P_0^{m+n} T_{0,\sigma(1\ldots N)}(\l)P_0^{m+n}
    R^\sigma_{1\ldots N}. \label{eq:T-tilde-1}
\end{eqnarray}
Consider the prefactor of $R^\sigma_{1\ldots N}$. We have
\begin{eqnarray}
 &&\prod_{j=1}^{N-k}P_{\sigma(j)}^{\alpha_{\sigma(j)}}
   \prod_{j=N-k+1}^N
    P_{\sigma(j)}^{{m+n}}
    P_0^{m+n} T_{0,\sigma(1\ldots N)}(\l)P_0^{m+n}\nonumber\\
 &=&\prod_{j=1}^{N-k}P_{\sigma(j)}^{\alpha_{\sigma(j)}}
   \prod_{j=N-k+1}^N\left(
   R_{0\,\sigma(j)}\right)^{{m+n}\ {m+n}}_{{m+n}\ {m+n}}
   P_0^{m+n} T_{0,\sigma(1\ldots N-k)}(\l)P_0^{m+n}
   \prod_{j=N-k+1}^N P_{\sigma(j)}^{{m+n}} \nonumber\\
 &=&
   \prod_{j=1}^{N-k}P_{\sigma(j)}^{\alpha_{\sigma(j)}}
   P_0^{m+n} T_{0,\sigma(1\ldots N-k)}(\l)P_0^{m+n}
   \prod_{j=N-k+1}^N P_{\sigma(j)}^{{m+n}} \nonumber\\
 &=&
   \prod_{i=1}^{N-k}\left(R_{0\,\sigma(i)}\right)
            ^{{m+n}\alpha_{\sigma(i)}}_{{m+n}\alpha_{\sigma(i)}}
   \prod_{j=1}^{N-k}P_{\sigma(j)}^{\alpha_{\sigma(j)}}
   \prod_{j=N-k+1}^N P_{\sigma(j)}^{{m+n}}\,\,P^{m+n}_0\nonumber\\
 &=&
   \prod_{i=1}^{N-k}a_{0\,\sigma(i)}
   \prod_{j=1}^{N-k}P_{\sigma(j)}^{\alpha_{\sigma(j)}}
   \prod_{j=N-k+1}^N P_{\sigma(j)}^{{m+n}}\,\,P^{m+n}_0,  \label{eq:T-tilde-2}
\end{eqnarray}
where $a_{0i}=a(\l,\x_i)$. Substituting (\ref{eq:T-tilde-2}) into
(\ref{eq:T-tilde-1}), we have
\begin{eqnarray}
F_{1\ldots N}D(\l)=\otimes_{i=1}^N
  \mbox{diag}\left(a_{0i},\ldots,a_{0i},1\right)_{(i)}
 F_{1\ldots N}.
\end{eqnarray}

\vskip 0.1in

By means of the expressions of the generators of $U_q(gl(m|n))$ in
theorem 3, combining with  the Theorem 1 and Proposition 3, we
have

\vskip 0.1in

\begin{Theorem}
In the $F$-basis the creation operators $C_{m+n-l}(\l)$ and
annihilation operators  $B_{m+n-l}(\l)$, $(l=1,\ldots,m+n-1)$, are
given by \bea
&&\tilde{C}_{m+n-l}(\l)=\lt(q^{-(-1)^{[m+n]}}\tilde{E}_{m+n-l,m+n}
\tilde{D}(\l)-\tilde{D}(\l)\tilde{E}_{m+n-l,m+n}\rt)
q^{-\sum_{k=1}^{l}H_{m+n-k}}\no\\
&&\quad\quad\mbox{}\hspace{-0.1truecm}-\hspace{-0.1truecm}
\sum_{\a=1}^{l-1}(1\hspace{-0.1truecm}-\hspace{-0.1truecm}
q^{-2(-1)^{[m+n-\a]}})\tilde{C}_{m+n-\a}(\l)
\tilde{E}_{m+n-l,m+n-\a} q^{-\sum_{k=\a+1}^{l}H_{m+n-k}}.
\label{Recursive1}\\[2mm]
%
&&\t B_{m+n-l}(\l)=
(-1)^{[m+n]+[m+n-l]}q^{\sum_{k=1}^lH_{m+n-k}}\nonumber\\
&&\quad\quad \mbox{}\times
 \lt(q^{(-1)^{[m+n]}}\t D(\l)\t E_{m+n,m+n-l}
  -\t E_{m+n,m+n-l}\t D(\l)\rt)
  \nonumber\\ &&\quad\mbox{}-
  \sum_{\a=1}^{l-1}(-1)^{([m+n]+[m+n-l])([m+n-l]+[m+n-l+\a])}
  (1-q^{2(-1)^{[m+n-l+\a]}})\no\\
&&\quad\quad\times
  q^{\sum_{k=l-\a+1}^{l}H_{m+n-k}}
  \t E_{m+n-l+\a,m+n-l}\t B_{m+n-l+\a}(\l),\label{Recursive2}
\end{eqnarray}
respectively.
\end{Theorem}

\vskip0.1in

\noindent For some special values  of $m$ and $n$, we have:\\
\begin{itemize}
\item For $m=2,n=0$,  i.e. the $U_q(gl(2|0))$ case:
 \bea
&& \tilde C_1(\l)
  =-\sum_{i=1}^Nb^-_{0i}
     E^{1,2}_{(i)}\otimes_{j\ne i}\mbox{diag}
     \left(2a_{0j}\cosh\eta,c_{0j}(2a_{ij}\cosh\eta)^{-1}\right)_{(j)}.
     \no\\ \label{eq:C-gl2}\\
&&\tilde B_1(\l)
 =-\sum_{i=1}^N b^+_{0i}E^{2,1}_{(i)}\otimes_{j\ne i}
  \mbox{diag}\left(a_{0j}(2a_{ji}\cosh\eta)^{-1},
     2c_{0j}\cosh\eta\right)_{(j)}, \no\\ \label{eq:B-gl2}
 \eea
\\
\item  For $m=2,n=1$, i.e. the $U_q(gl(2|1))$ case,
 \begin{eqnarray}
\tilde C_2(\l)
 &=&\sum_{i=1}^Nb_{0i}^{-}E^{2,3}_{(i)}
 \otimes_{j\ne i}\mbox{diag}\left(a_{0j},2a_{0j}\cosh\eta,1\right)_{(j)},
 \label{eq:C2-tilde}\\
\tilde B_2(\l)
 &=&-\sum_{i=1}^Nb_{0i}^{+}E^{3,2}_{(i)}
 \otimes_{j\ne i}\mbox{diag}\left(a_{0j},a_{0j}(2a_{ji}\cosh\eta)^{-1},
     a_{ji}^{-1}\right)_{(j)},
 \label{eq:B2-tilde}\\
%
\tilde C_1(\l)
 &=&\sum_{i=1}^N b_{0i}^-
 E^{1,3}_{(i)}
 \otimes_{j\ne i}\mbox{diag}\left(2a_{0j}\cosh\eta,
 a_{0j}(a_{ij})^{-1},1\right)_{(j)}\nonumber\\
 && \mbox{}+\sum_{i\ne j=1}^N
  {a_{0i}b^-_{0j}(2 a_{ij}\,\sinh\eta+b^-_{ij})
   \over a_{ij}}
   E^{1,2}_{(i)}\otimes E^{2,3}_{(j)}
  \nonumber\\&&\quad
  \otimes_{k\ne i,j}\left(2a_{0k}\cosh\eta,a_{0k}(a_{ik})^{-1},1
 \right)_{(k)}. \label{eq:C1-tilde}\\
\tilde B_1(\l)
 &=&-\sum_{i=1}^N b_{0i}^+
 E^{3,1}_{(i)}
 \otimes_{j\ne i}\mbox{diag}\left(a_{0j}(2a_{ji}\cosh\eta)^{-1},
 a_{0j}(a_{ji})^{-1},a_{ji}^{-1}\right)_{(j)}\nonumber\\
 && \mbox{}+\sum_{i\ne j=1}^N
  {a_{0j}b^+_{0i}(2a_{ji}\,\sinh\eta
   -b^+_{ji})\over a_{ji}}
  E^{3,2}_{(i)}\otimes E^{2,1}_{(j)}
  \nonumber\\&&\quad
  \otimes_{k\ne
  i,j}\left(a_{0k}(2a_{kj}\cosh\eta)^{-1},a_{0k}(a_{ki})^{-1},a_{ki}
 \right)_{(k)}. \nonumber\\ \label{eq:B1-tilde}
\end{eqnarray}
\end{itemize}
Here $x_{0k}\equiv x(\l,\x)$ with $x_{ij}$ stands for $a_{ij},
b^{\pm}_{ij}$.


Thus, one sees that working in the $F$-basis, the creation and
annihilation operators take completely symmetric forms, e.g.
(\ref{eq:C-gl2})-(\ref{eq:B1-tilde}).


\sect{The $q$-deformed supersymmetry $t$-$J$ model}

In this section, we will construct the determinant representation
of the correlation functions for the $q$-deformed supersymmetric
$t$-$J$ model, generalizing the results in \cite{zsy0511} for the
supersymmetric $t$-$J$ model.

\subsection{Bethe states}

In the framework of the algebraic ansatz, the Bethe states
(eigenstates) are constructed by acting the creation operators
$C_i(\l)$ (or the annihilation operators $B_i(\l)$) on the
pseudo-vacuum state.
\begin{Definition}
Let $|0\rangle$ be the pseudo-vacuum state of the $N$-fold quantum
tensor space $V^{\otimes N}$, and $|0\rangle^{(1)}$ be the
pseudo-nested-vacuum state of the $n_1$-fold nested quantum tensor
space $(V^{(1)})^{\otimes n_1}$, i.e.,
\begin{equation}
 |0\rangle=\otimes_{i=1}^N\left(\begin{array}{c} 0\\0\\1\end{array}
 \right)_{(i)}, \quad\quad
 |0\rangle^{(1)}=\otimes_{j=1}^{n_1}\left(\begin{array}{c}
 0\\1\end{array} \right)_{(j)}.
 \label{de:vacuum-tj}
\end{equation}
The Bethe state of the $q$-deformed supersymmetric $t$-$J$ model
is then defined by
\begin{eqnarray}
|\Omega_N(\{\l_j\})\rangle=\sum_{d_1\ldots d_{n_1}}
  (\Omega^{(1)}_{n_1})^{d_1\ldots d_{n_1}}C_{d_1}(\l_1)\ldots
  C_{d_{n_1}}(\l_{n_1})|0\rangle \,\,(\l_1\ne\ldots\ne\l_{n_1}),
 \label{de:Omega}
\end{eqnarray}
where $d_i=1,2$, $(\Omega^{(1)}_{n_1})^{d_1\ldots d_{n_1}}$ is a
component of the nested Bethe state $|\O\rangle^{(1)}$ via
\begin{equation}
|\Omega_{n_1}(\{\l_j^{(1)}\})\rangle^{(1)}=C_1^{(1)}(\l^{(1)}_1)
 \cdots C_1^{(1)}(\l^{(1)}_{n_2})|0\rangle^{(1)}
 \,\,\,\,\,(\l^{(1)}_1\ne\ldots\ne \l^{(1)}_{n_2}),
 \label{de:Bethe-vector-nested}
\end{equation}
and $C^{(1)}$, the creation operator of the nested $U_q(gl(2))$
system, is the lower-triangular entry of the nested monodromy
matrix $T^{(1)}$
\begin{eqnarray}
T^{(1)}(\l^{(1)})&=&r_{0n_1}(\l^{(1)}-\l_{n_1})r_{0n_1-1}(\l^{(1)}-\l_{n_1-1})
 \ldots r_{01}(\l^{(1)}-\l_1) \label{de:T-nested}
\nonumber\\
&\equiv& \left(\begin{array}{ccc}
 A^{(1)}(\l^{(1)})& B^{(1)}_1(\l^{(1)})\\
 C^{(1)}_1(\l^{(1)})& D^{(1)}(\l^{(1)})\end{array}\right)_{(0)}
\end{eqnarray}
with
$r_{12}(\l_1,\l_2)\equiv r_{12}(\l_1-\l_2)=
 c_{12}(e_{1,1}\otimes e_{1,1}+e_{2,2}\otimes e_{2,2})
  +a_{12}(e_{1,1}\otimes e_{2,2}+e_{2,2}\otimes e_{1,1}
  -b^-_{12}e_{2,1}\otimes e_{1,2}-b^+_{12}e_{1,2}\otimes e_{2,1} .$
\end{Definition}
Similarly, we can also define the dual Bethe state
$\langle\Omega_N|$.
\begin{Definition}
With the help of the dual pseudo-vacuum state $\langle 0|$ and the
dual pseudo-nested-vacuum state $\langle 0|^{(1)}$, the dual Bethe
state is defined by
\begin{eqnarray}
\langle
\O_N(\{\m_j\})|=\sum_{f_{n_1},\ldots,f_1}(\O^{(1)})^{f_{n_1}\ldots
f_1}
 \langle0| B_{f_{n_1}}(\m_{n_1})\ldots B_{f_1}(\m_1)
 \,(\m_{n_1}\ne\ldots\ne \m_1),\label{de:dual-state}
\end{eqnarray}
where $(\O^{(1)})^{f_{n_1}\ldots f_1}$ is a component of the dual
nested Bethe state $\langle\O|^{(1)}$
\begin{eqnarray}
 \langle\O_{n_1}(\{\m_j^{(1)}\})|^{(1)}=\langle 0|^{(1)}B_1^{(1)}(\m_{n_2}^{(1)})
  \ldots B_1^{(1)}(\m_1^{(1)})\quad\quad (\m_{n_2}^{(1)}\ne\ldots\ne\m_1^{(1)}).
\end{eqnarray}
\end{Definition}

The diagonalization of the transfer matrix $t(\l)$ leads to the
following theorem \cite{Essler9207}:
\begin{Theorem}
The Bethe states $|\O_N(\{\l_j\})\rangle$ defined by
(\ref{de:Omega}) are eigenstates of the transfer matrix $t(\l)$ if
the spectral parameters $\l_j$ $(j=1,\ldots, n_1)$ satisfy the
Bethe ansatz equations (BAE)
\begin{eqnarray}
&&\prod_{k=1}^N a(\l_j,\x_k)
\prod_{l=1}^{n_2}a^{-1}(\l_j,\l_l^{(1)})=1
 \quad\quad (j=1,\ldots,n)
\label{eq:BAE}
\end{eqnarray}
and the nested Bethe ansatz equations (NBAE)
\begin{eqnarray}
&&\prod_{j=1}^{n_1} a(\l_j,\l^{(1)}_l)\prod_{k=1,\ne l}^{n_2}
  {a(\l_l^{(1)},\l_k^{(1)})\over a(\l_k^{(1)},\l_l^{(1)})}=1
\quad\quad (l=1,\ldots,{n_2}). \label{eq:BAE-nested}
\end{eqnarray}
The eigenvalues $\Lambda(\l,\{\l_k\},\{\l^{(1)}_j\})$ of the
transfer matrix $t(\l)$ are given by
\begin{eqnarray}
&&\Lambda(\l,\{\l_k\},\{\l^{(1)}_j\})\nonumber\\
 &&=\prod_{i=1}^Na(\l,\x_i)
    \prod_{j=1}^{n_1}{1\over a(\l,\l_j)}\Lambda^{(1)}(\l)
    +\prod_{j=1}^{n_1}{1\over a(\l_j,\l)},
    \label{eq:eigenvalue}
\end{eqnarray}
where $\Lambda^{(1)}(\l)$ is the eigenvalues of the nested
transfer matrix $t^{(1)}(\l)=\str_0T^{(1)}(\l)$
\begin{eqnarray}
 \Lambda^{(1)}(\l)=-\prod_{j=1}^{n_1}{a(\l,\l_j)\over a(\l_j,\l)}
 \prod_{k=1}^{n_2}{1\over a(\l,\l^{(1)}_k)}
 -\prod_{j=1}^{n_1} a(\l,\l_j)
 \prod_{k=1}^{n_2}{1\over a(\l^{(1)},\l)}.   \label{eq:eigenvalue-nested}
\end{eqnarray}
\end{Theorem}
One easily checks that this theorem also holds for the dual Bethe
state $\langle \O_N(\{\m_j\})|$ defined by (\ref{de:dual-state})
if we change the spectral parameters $\l_j$ and $\l_j^{(1)}$ in
(\ref{eq:BAE})-(\ref{eq:eigenvalue-nested}) to $\m_j$ and
$\m_j^{(1)}$, respectively.

\vskip12pt

Let $\s=\s_{i_1}\ldots\s_{i_k}$ be any element of the  permutation
group ${\cal S}_{n_1}$ with $\s_{i_j}$ be elementary permutations
$\s_{i_j}(i_j,i_{j}+1)=(i_j+1,i_j)$.  We define the following
exchange operator $\hat f_\sigma=\hat f_{\sigma_{i_1}}\ldots \hat
f_{\sigma_{i_k}}$ for the Bethe state $|\Omega_N(\{\l_j\})\rangle$
of the  $q$-deformed supersymmetric $t$-$J$ model,
$ \hat f_\sigma |\Omega_N(\{\l_j\})\rangle
 =|\Omega_N(\{\l_{\s(j)}\})\rangle.$
Then one may prove \cite{zsy04,zsy0503}, under the action of the
exchange operator, we have the following proposition:

\begin{Proposition}
The Bethe state $|\Omega_N(\{\l_j\})\rangle$ of the  $q$-deformed
supersymmetric $t$-$J$ model satisfies the following exchange
symmetry
\begin{eqnarray}
  \hat f_{\sigma} |\Omega_N(\{\l_j\})\rangle
 &=&{1\over c^\sigma_{1\ldots \a}}
 |\Omega_N(\{\l_j\})\rangle, \label{eq:Exchange}
\end{eqnarray}
where $c^\sigma_{1\ldots n_1}$ has the decomposition law
$ c^{\sigma'\sigma}_{1\ldots n_1}
 =c^{\sigma}_{\sigma'(1\ldots n_1)}
  c^{\sigma'}_{1\ldots n_1}\label{eq:c-cc},$
and  $c^{\sigma_i}_{1\ldots n_1}=c_{i\ i+1}\equiv
c(\l_i,\l_{i+1})$ for an elementary permutation $\sigma_i$.

\end{Proposition}

For the non-super case, the element $c$ of the $R$-matrix
(\ref{R12}) tends to 1 so the relation (\ref{eq:Exchange}) of the
exchange symmetry changes to $ \hat f_{\sigma}
|\P_N(\{\l_j\})\rangle =|\P_N(\{\l_j\})\rangle,$ where
$|\P\rangle$ is the Bethe state of any integrable $U_q(gl(n))$
system. The exchange symmetry for the non-super case was first
proposed by \cite{Vega89,Zam80}.


\subsection{Symmetric representations of the Bethe state}

Acting the associated $F$-matrix on the pseudo-vacuum state
$|0\rangle$, one finds that the pseudo-vacuum state is invariant.
It is due to the fact that only the term with all roots equal to 3
will produce  non-zero results. Therefore, the Bethe state
(\ref{de:Omega}) in the $F$-basis $ |\tilde\O_N(\{\l_j\})\rangle
 \equiv F_{1\ldots N}|\O_N(\{\l_j\})\rangle$ can be written as
\begin{eqnarray}
   |\tilde\O_N(\{\l_j\})\rangle
 = \sum_{d_1\ldots d_{n_1}}
  (\Omega^{(1)}_{n_1})^{d_1\ldots d_{n_1}}\tilde C_{d_1}(\l_1)\ldots \tilde
  C_{d_{n_1}}(\l_{n_1})|0\rangle. \label{eq:Omega-F}
\end{eqnarray}
Without loss of generality, we will only concentrate on the Bethe
state with the {\bf quantum number $p$} which indicates the number
of $d_i=2$, and will use the notation
$|\tilde\O_N(\{\l_j\}_{(p,n_1)})\rangle$ with the subscript pair
$(p,n_1)$ to denote a Bethe state which has quantum number $p$ and
has $n$ spectral parameters.
\begin{Proposition}
The Bethe state of the  $q$-deformed supersymmetric t-J model can
be represented in the $F$-basis  by
\begin{eqnarray}
 && |\tilde\O_N(\{\l_j\}_{(p,n_1)})\rangle
=\sum_{\s\in{\cal S}_N}Y_R(\{\l_{\s(i)}\},\{\l_{\s(j)}^{(1)}\})
 \nonumber\\ && \times
 {\tilde C}_{2}(\l_{\sigma(1)})\ldots {\tilde C}_{2}(\l_{\sigma(p)})
 {\tilde C}_{1}(\l_{\sigma(p +1)})\ldots
 {\tilde C}_{1}(\l_{\sigma(n_1)})\,|0\rangle \label{eq:Omega-YR} \\
 &=&\sum_{\s\in{\cal S}_N}Y_R(\{\l_{\s(i)}\},\{\l_{\s(j)}^{(1)}\})
     \sum_{i_1<\ldots<i_{p}}\sum_{i_{p+1}<\ldots<i_{n_1}}
   (2\cosh\eta)^{{p(p-1)+(n_1-p)(n_1-p-1)\over 2}}
  \nonumber\\ &&\times
  \prod_{l=1}^{p}~\prod_{k=p +1}^{n_1}a(\l_{\sigma(l)},\x_{i_{k}})
   \mbox{det}\,{\cal B}^-_p(\l_{\s(1)},\ldots,\l_{\s(p)};
   \x_{i_1},\ldots,\x_{i_p})
 \nonumber\\ &&\times
 \mbox{det}\,{\cal B}^-_{n-p}(\l_{\s(p+1)},\ldots,\l_{\s(n_1)};
   \x_{i_{p+1}},\ldots,\x_{i_{n_1}})
   \overrightarrow{\prod_{j=1}^{p}}E^{23}_{(i_j)}
   \overrightarrow{\prod_{j=p+1}^{n_1}}E^{13}_{(i_j)}|0\rangle
   \label{eq:state-proposition1}
\end{eqnarray}
with the sets
$\{i_1,\ldots,i_p\}\cap\{i_{p+1},\ldots,i_{n_1}\}=\varnothing$ and
the prefactor $Y_R$ being
\begin{eqnarray}
 &&Y_R(\{\l_{\s(i)}\},\{\l_{\s(j)}^{(1)}\} )
={1\over p!(n_1-p)!} c^{\sigma}_{1\ldots n_1}
   \prod_{k=p +1}^{n_1}~\prod_{l=1}^{p}
    \left(-{2a(\l_{\sigma(k)},\l_{\sigma(l)})\cosh\eta\over
     a(\l_{\sigma(l)},\l_{\sigma(k)})}\right)
    \nonumber\\ && \times
    B^*_{n-p}\left(\l_{p +1}^{(1)},\ldots,\l_{n_1}^{(1)}|
     \l_{\sigma(p +1)},\ldots,\l_{\sigma(n_1)}\right).
\end{eqnarray}
Here $c^\sigma_{1,\ldots,n_1}$ has the decomposition law
 $c^{\sigma'\sigma}_{1\ldots n_1}
 =c^{\sigma}_{\sigma'(1\ldots n_1)}
  c^{\sigma'}_{1\ldots n_1}$
with $c^{\sigma_i}_{1\ldots n_1}=c_{i\ i+1}\equiv
c(\l_i,\l_{i+1})$ for an elementary permutation $\sigma_i$, the
$c$-number $B_{p}^*$ is given by
\begin{eqnarray*}
&&B^*_{p}\left(\l_1^{(1)},\ldots,\l_{p}^{(1)}|
                \l_{1},\ldots,\l_{p}\right)
 =\sum_{\sigma\in{\cal S}_{p}}\prod_{k=1}^{p}
  \left(-b^-(\l_k^{(1)},\l_{\sigma(k)})\right)
\nonumber\\ &&\times
  \prod_{j\ne \sigma(k),\ldots,\sigma(p)}
  {c(\l_k^{(1)},\l_{j})\over 2a(\l_{\s(k)},\l_{j})\cosh\eta}
  \prod_{l=k+1}^{p} 2a(\l_k^{(1)},\l_{\sigma(l)})\cosh\eta,
\end{eqnarray*}
and the elements of the $n_1\times n_1$ matrix ${\cal
B}^\pm_{n_1}(\{\l_i\};\{\x_{j}\})$ are
\begin{equation}
({\cal B}^\pm_{n_1})_{\alpha\beta}
 =b^\pm(\l_\alpha,\x_\beta)
  \prod_{\gamma=1}^{\alpha-1}a(\l_\gamma,\x_\beta).
\end{equation}
\end{Proposition}
In (\ref{eq:state-proposition1}), we have used the convention
$\overrightarrow{\prod_{i=1}^{n_1}}f_i\equiv f_1\ldots f_{n_1}$.
For our later use, we also introduce the notation
$\overleftarrow{\prod_{i=1}^{n_1}}f_i\equiv f_{n_1}\ldots f_1$.

\vskip12pt

\noindent {\it Proof}. The proof is similar to that for the
supersymmetric $t$-$J$ model \cite{zsy0503}. Define
$|\tilde\Omega_{n_1}^{(1)}(\{\l_j^{(1)}\})\rangle
 \equiv F^{(1)}_{1\ldots n_1}
|\Omega_{n_1}^{(1)}(\{\l_j^{(1)}\})\rangle$, where
$F^{(1)}_{1\ldots n_1}$ is the $F$-matrix (\ref{de:F}) with
$m=2,n=0$. Then substituting the expression $\t C_1$
(\ref{eq:C-gl2}) into the nested Bethe state
(\ref{de:Bethe-vector-nested}), we have
\begin{eqnarray}
 && |\tilde\Omega_{n_1}^{(1)}(\{\l_j^{(1)}\})\rangle
   =s(c)\tilde{C}^{(1)}(\l_1^{(1)})\ldots \tilde{C}^{(1)}(\l_{n_2}^{(1)})
   \,|0\rangle^{(1)}\nonumber\\
&=&s(c)\sum_{i_1<\ldots< i_{n_2}}
B_{n_2}^{(1)}(\l_1^{(1)},\ldots,\l_{n_2}^{(1)}|\l_{i_1},\ldots,\l_{i_{n_2}})
E_{(i_1)}^{1,2}\ldots E_{(i_{n_2})}^{1,2}\,|0\rangle^{(1)}\;,
\label{Psi_2}
\end{eqnarray}
where $s(c)=\prod_{i<j}(1+c_{ij}),$ and
\begin{eqnarray}
&&
B^{(1)}_{n_2}(\l_1^{(1)},\ldots,\l_{n_2}^{(1)}|\l_{1},\ldots,\l_{n_2})
 =\sum_{\sigma\in {\cal S}_{n_2}}\prod_{k=1}^{n_2}
 \left(-b^-(\l_k^{(1)},\l_{\sigma(k)})\right)
\nonumber\\ && \times
  \prod^{n_1}_{j\ne \sigma(k),\ldots,\sigma(\b)}
  {c(\l_k^{(1)},\l_j)\over 2a(\l_{\sigma(k)},\l_j)\cosh\eta}
 \prod_{l=k+1}^{n_2}
 {2a(\l_k^{(1)},\l_{\sigma(l)})\cosh\eta}. \label{eq:B1}
\end{eqnarray}

Now we study the Bethe state (\ref{de:Omega}) of the quantum
supersymmetric $t$-$J$ model.
\begin{eqnarray}
&&|\t\O_N(\{\l_j\})\rangle=\sum_{d_1\ldots d_{n_1}}
  (\Omega^{(1)}_{n_1})^{d_1\ldots d_{n_1}}\t C_{d_1}(\l_1)\ldots
  \t C_{d_{n_1}}(\l_{n_1})|0\rangle\nonumber\\
&=&(\Omega^{(1)}_{n_1})^{1\ldots 12\ldots 2}
  \t C_{1}(\l_{p+1})\ldots \t C_1(\l_{n_1})
  \t C_{2}(\l_{1})\ldots \t C_2(\l_p)|0\rangle
 +\mbox{other terms}\no\\
&=&{1\over p!(n_1-p)!}\sum_{\sigma \in {\cal S}_{n_1}}
     c^{\sigma}_{1\ldots n_1}
    (\Omega^{(1),\sigma}_{n_1})^{11\ldots 12\ldots 2}
    \prod_{k=p+1}^{n_1}~\prod_{l=1}^{p}
    \left(-{1\over a(\l_{\sigma(l)},\l_{\sigma(k)})}\right)
    \nonumber\\
&&\times\;{\tilde C}_{2}(\l_{\sigma(1)})\ldots {\tilde
C}_{2}(\l_{\sigma(p)}) {\tilde C}_{1}(\l_{\sigma(p+1)})\ldots
{\tilde C}_{1}(\l_{\sigma(n_1)})\,|0\rangle\, , \label{eq:Phi-phi}
\end{eqnarray}
where we have used the proposition 4 and the commutation relation
\begin{eqnarray}
 \tilde C_i(\m)\tilde C_j(\l)
 &=&-{1\over a(\l,\m)}\tilde C_j(\l)\tilde C_i(\m)
    +{b(\l,\m)\over a(\l,\m)}\tilde C_j(\m)\tilde C_i(\l).
    \label{eq:commu-CC-F}
\end{eqnarray}
Considering the $1\ldots 12\ldots 2$ component of the nested Bethe
state $|\t\Omega_{n_1}^{(1)}(\{\l_j^{(1)}\})\rangle$, one easily
proves \cite{zsy0503} the relation
$ (\tilde\Omega^{(1)}_{n_1})^{11\ldots 12\ldots 2}
 =t(c)(\Omega^{(1)}_{n_1})^{11\ldots 12\ldots 2}$
with  the scalar factor $t(c)=\prod_{j>i=1}^{p}(1+\bar{c}_{ij})
    \prod_{j>i=p_1+1}^{n_1}(1+\bar{c}_{ij}),$ and $\bar{c}_{ij}=c(\l_i,\l_j).$
Then substituting the expressions $\t C_1$ (\ref{eq:C1-tilde}) and
$\t C_2$  (\ref{eq:C2-tilde}) into (\ref{eq:Phi-phi}), one obtains
(\ref{eq:state-proposition1}). Therefore we have proved the
proposition 4. ~~~~~~~~~~~~~~~~~~~~~~~
~~~~~~~~~~~~~~~~~~~~~~~~~~~~~~~~~~~~~~~~~~~~~~~~~~~~~~$\Box$

\vskip12pt

By a similar procedure, one may prove the following proposition
for the dual Bethe state $\langle\t\O_N(\{\m_j\}_{(p,n_1)})|$
(\ref{de:dual-state}):
\begin{Proposition}
The dual Bethe state $\langle\t\O_N(\{\m_j\}_{(p,n_1)})|$ of the
$\gl$ supersymmetric t-J model can be represented by
\begin{eqnarray}
 && \langle\t\O_N(\{\m_j\}_{(p,n_1)}|
 =\sum_{\s\in{\cal S}_N}Y_L(\{\m_{\s(i)}\},\{\m_{\s(j)}^{(1)}\})
 \langle0|
 {\tilde B}_{1}(\m_{\sigma(n_1)})\ldots {\tilde B}_{1}(\m_{\sigma(p+1)})
 \nonumber\\&&\times
 {\tilde B}_{2}(\m_{\sigma(p)})\ldots
 {\tilde B}_{2}(\m_{\sigma(1)})\, ~~~~~~\label{eq:Omega-YL} \\
 &=&\sum_{\s\in{\cal S}_N}Y_L(\{\m_{\s(i)}\},\{\m_{\s(j)}^{(1)}\})
     \sum_{i_1<\ldots<i_{p}}\sum_{i_{p+1}<\ldots<i_{n_1}}
   (-1)^{n_1} (2\cosh\eta)^{-{p(p-1)+(n_1-p)(n_1-p-1)\over 2}}
  \nonumber\\ &&\times
  \prod_{l=1}^{p}~\prod_{k=p +1}^{n_1}a(\m_{\sigma(l)},\x_{i_{k}})
  \prod_{l=1}^p\prod_{k\ne i_l,i_{p+1},\ldots,i_{n_1}}^N a^{-1}(\x_{k},\x_{i_l})
  \prod_{l=p+1}^{n_1}\prod_{k=1,\ne i_l}^N a^{-1}(\x_{k},\x_{i_l})
  \nonumber\\ &&\times
   \mbox{det}\,{\cal B}^+_p(\m_{\s(1)},\ldots,\m_{\s(p)};
   \x_{i_1},\ldots,\x_{i_p})
 \nonumber\\ &&\times
 \mbox{det}\,{\cal B}^+_{n_1-p}(\m_{\s(p+1)},\ldots,\m_{\s(n_1)};
   \x_{i_{p+1}},\ldots,\x_{i_{n_1}})
   \langle0|\overleftarrow{\prod_{j=p+1}^{n_1}}E^{31}_{(i_j)}
   \overleftarrow{\prod_{j=1}^{p}}E^{32}_{(i_j)},
\end{eqnarray}
where the prefactor $Y_L$ is
\begin{eqnarray}
 &&Y_L(\{\m_{\s(i)}\},\{\m_{\s(j)}^{(1)}\} ) \nonumber\\
 &&={1\over p!(n_1-p)!}
  c^{\sigma}_{1\ldots n_1}
  B^{**}_{n-p}\left(\m_{p +1}^{(1)},\ldots,\m_{n_1}^{(1)}|
     \m_{\sigma(p +1)},\ldots,\m_{\sigma(n_1)}\right)
  \nonumber\\ &&\times
    \prod_{k=p +1}^{n_1}~\prod_{l=1}^{p}
    \left(-{2a(\m_{\sigma(k)},\m_{\sigma(l)})\cosh\eta\over
     a(\m_{\sigma(l)},\m_{\sigma(k)})}\right),\nonumber\\
\end{eqnarray}
and the $c$-number $B_{p}^{**}$ is given by
\begin{eqnarray*}
&&B^{**}_{p}\left(\m_1^{(1)},\ldots,\m_{p}^{(1)}|
                \m_{1},\ldots,\m_{p}\right)
\nonumber\\
 &&=\sum_{\sigma\in{\cal S}_{p}}\prod_{k=1}^{p}
  b^+(\m_k^{(1)},\m_{\sigma(k)})
  \prod_{j\ne \sigma(k),\ldots,\sigma(p)}c(\m_k^{(1)},\m_{j})
  \prod_{l=k+1}^{p}
  {a(\m_k^{(1)},\m_{\sigma(l)})\over
  2a(\m_{\s(l)},\m_{\s(k)})\cosh\eta}\,.
\end{eqnarray*}
\end{Proposition}


\subsection{Determinant representation of the scalar product }

The scalar product of the Bethe states with a given quantum number
$p$ is defined by
\begin{eqnarray}
 \mathbb{P}_{n_1}(\{\m_i\}_{(p,n_1)},\{\l_j\}_{(p,n_1)})&=&
 \langle\O_N(\{\m_j\}_{(p,n_1)})|\O_N(\{\l_j\}_{(p,n_1)})\rangle.
 \label{de:P_{n_1}}
\end{eqnarray}
The invariant property of the pseudo-vacuum state under the
$F$-transformation, i.e. $F_{1\ldots N}|0\rangle=|0\rangle$ and
$\langle0|F^{-1}_{1\ldots N}=\langle0|$, implies that in the
$F$-basis, the scalar product $\mathbb{P}_{n_1}$ is
\begin{eqnarray}
 &&\mathbb{P}_{n_1}(\{\m_i\}_{(p,n_1)},\{\l_j\}_{(p,n_1)})
  =
 \langle\tilde \O_N(\{\m_j\}_{(p,n_1)})|
 \tilde\O_N(\{\l_j\}_{(p,n_1)})\rangle
 \nonumber\\ &&=\sum_{\s',\s}
 Y_L(\{\m_{\sigma'(j)}\},\{\m^{(1)}_{\sigma'(k)}\})
 Y_R(\{\l_{\sigma(j)}\},\{\l^{(1)}_{\sigma(k)}\})
 \nonumber\\&& \quad\times
 \langle 0|\tilde B_1(\m_{\s'(n_1)})\ldots \tilde B_1(\m_{\s'(p+1)})
 \tilde B_2(\m_{\s'(p)})\ldots \tilde B_2(\m_{\s'(1)})
  \nonumber\\&&\quad \times
  \tilde C_{2}(\l_{\s(1)})\ldots \tilde C_{2}(\l_{\s(p)})
  \tilde C_{1}(\l_{\s(p+1)})\ldots \tilde C_{1}(\l_{\s(n_1)})|0\rangle.
  \label{de:S_{n_1}}
\end{eqnarray}

To compute the scalar product, following
\cite{Kitanine98,zsy0511}, we introduce the following intermediate
functions $G^{(m)}$ (in this section, the integer $m$ is the index
of the intermediate function)
\begin{eqnarray}
&&G^{(m)}(\{\l_k\}_{(p,n_1)},\m_1,\ldots,\m_m,i_{m+1},\ldots,i_{n_1})\nonumber\\
 &&=\left\{\begin{array}{l}
 \langle0|\overleftarrow{\prod_{k=p+1}^{n_1}}E_{(i_k)}^{31}
    \overleftarrow{\prod_{k=m+1}^{p}}E_{(i_k)}^{32}
 \tilde B_2(\m_m)\ldots \tilde B_2(\m_1) \\ \quad\quad \times
  \tilde C_2(\l_1)\ldots \tilde C_2(\l_{p})
  \tilde C_1(\l_{p+1})\ldots \tilde C_1(\l_{n_1})|0>\quad
    \mbox{ for }m\leq p,\\[3mm]
\langle0|\overleftarrow{\prod_{k=m+1}^{n_1}}E_{(i_k)}^{31}
 \tilde B_1(\m_m)\ldots \tilde B_1(\m_{p+1})
 \tilde B_2(\m_{p}) \tilde B_2(\m_1)
  \\ \quad\quad \times
  \tilde C_2(\l_1)\ldots \tilde C_2(\l_{p})
  \tilde C_1(\l_{p+1})\ldots \tilde C_1(\l_{n_1})|0>\quad
    \mbox{ for }m\geq p+1.
  \end{array}\right.
  \label{de:G-m}
\end{eqnarray}
where the lower indices of $E^{32}_{(i_k)}$ and $E^{31}_{(i_k)}$
satisfy the relations $i_{m+1}<\ldots<i_{p}$,
$i_{p+1}<\ldots<i_{n_1}$ and
$\{i_1,\ldots,i_p\}\cap\{i_{p+1},\ldots,i_{n_1}\}=\varnothing$.
Thus, the scalar product can be rewritten as
\begin{eqnarray}
&&\mathbb{P}_{n_1}(\{\m_i\}_{(p,n_1)},\{\l_j\}_{(p,n_1)})
=\sum_{\s,\s'}Y_L(\{\m_{\sigma'(j)}\},\{\m^{(1)}_{\sigma'(k)}\})
\nonumber\\
 &&\times
 Y_R(\{\l_{\sigma(j)}\},\{\l^{(1)}_{\sigma(k)}\})\,\
 G^{(n_1)}(\{\l_{\s(j)}\}_{(p,n_1)},\{\m_{\s'(k)}\}_{(p,n_1)}).
 \label{eq:Sn-Gn}
\end{eqnarray}

We now compute $G^{(m)}$ for $m\leq p$ and $m\geq p+1$ separately.
The procedure is similar to that for the supersymmetric $t$-$J$
model \cite{zsy0511}.

\begin{itemize}
\item{$1\leq m\leq p$}
\end{itemize}

We first compute the function $G^{(m)}$ for $m\leq p$.

Inserting a complete set, (\ref{de:G-m}) becomes
\begin{eqnarray}
&&G^{(m)}(\{\l_k\}_{(p,n_1)},\m_1,\ldots,\m_{m},i_{m+1},\ldots,i_{n_1})
=\sum_{j\ne i_{m+1},\ldots,i_{n_1}}^N
\nonumber\\
 &&
     \langle0|\overleftarrow{\prod_{k=p+1}^{n_1}}E_{(i_k)}^{31}
    \overleftarrow{\prod_{k=m+1}^{p}}E_{(i_k)}^{32}
    \tilde B_2(\m_{m})
     \overrightarrow{\prod_{k=m+1}^{m+q}}E_{(i_k)}^{23}E_{(j)}^{23}
     \overrightarrow{\prod_{k=m+q+1}^{p}}E_{(i_k)}^{23}
     \overrightarrow{\prod_{k=p+1}^{n_1}}E_{(i_k)}^{13}|0\rangle \nonumber\\
 &&\times
 G^{(m-1)}(\{\l_k\}_{(p,n_1)},\m_1,\ldots,\m_{m-1},i_{m+1},
 \ldots,i_{m+q},j,i_{m+q+1}\ldots,i_{n_1}),
 \label{de:Gm-Gm-1}
\end{eqnarray}
where $(0\leq q\leq p-m).$ In view of (\ref{eq:B2-tilde}), we have
\begin{eqnarray}
&&\langle0|\overleftarrow{\prod_{k=p+1}^{n_1}}E_{(i_k)}^{31}
    \overleftarrow{\prod_{k=m+1}^{p}}E_{(i_k)}^{32}
    \tilde B_2(\m_{m})
     \overrightarrow{\prod_{k=m+1}^{m+q}}E_{(i_k)}^{23}E_{(j)}^{23}
     \overrightarrow{\prod_{k=m+q+1}^{p}}E_{(i_k)}^{23}
     \overrightarrow{\prod_{k=p+1}^{n_1}}E_{(i_k)}^{13}|0\rangle
\nonumber\\
 &=&-(-1)^{q}(2\cosh\eta)^{-(p-m)}\cdot\,
 b^+(\m_{m},\x_j)\prod_{l=m+1}^p a(\m_{m},\x_{i_l})
 \no\\ && \times
\prod_{l=p+1}^{n_1} a(\m_{m},\x_{i_l})
 \prod_{k\ne j,i_{p+1},\ldots,i_{n_1}}^Na^{-1}(\x_k,\x_j)
 .  \label{eq:B-expect}
\end{eqnarray}

Substituting the expressions of $\tilde C_1$ (\ref{eq:C1-tilde})
and $\tilde C_2$ (\ref{eq:C2-tilde}) into (\ref{de:G-m}), we
obtain $G^{(0)}$:
\begin{eqnarray}
&& G^{(0)}(\{\l_k\}_{(p,n_1)},i_1,\ldots,i_{n_1})
 =\langle0|\overleftarrow{\prod_{k=p+1}^{n_1}}E_{(i_k)}^{31}
    \overleftarrow{\prod_{k=1}^{p}}E_{(i_k)}^{32}
    \prod_{k=1}^{p}\tilde C_2(\l_k)
    \prod_{k=p+1}^{n_1}\tilde C_1(\l_k)|0> \nonumber\\
 &&=2^{{p(p-1)+(n_1-p)(n_1-p-1)\over 2}}
    \prod_{l=1}^{p}\prod_{k=p+1}^{n_1}
    a(\l_{l},\x_{i_k})
 \mbox{det} {\cal B}_{p}(\l_1,\ldots,\l_{p};
      \x_{i_1},\ldots,\x_{i_{p}})
 \nonumber\\ &&\quad\times
 \mbox{det} {\cal B}_{n_1-p}(\l_{p+1},\ldots,\l_{n_1};
      \x_{i_{p+1}},\ldots,\x_{i_{n_1}}). \label{eq:G-0}
\end{eqnarray}

We compute $G^{(m)}$ by using the recursion relation
(\ref{de:Gm-Gm-1}). One sees that there are two different
determinants in $G^{(0)}$, which are labelled by different $\l$'s
and $\x_{i_{k}}$'s. For  $m\leq p$ we only focus on the first
determinant, i.e. $\mbox{det}{\cal B}_{p}$.

To compute $G^{(1)}$, we substitute (\ref{eq:B-expect}) and
(\ref{eq:G-0}) into (\ref{de:Gm-Gm-1}) to obtain
\begin{eqnarray}
&& G^{(1)}(\{\l_k\}_{(p,n_1)},\m_1,i_2,\ldots,i_{n_1})\nonumber\\
& =&\sum^N_{j\ne
i_2,\ldots,i_{n_1}}\langle0|\overleftarrow{\prod_{k=p+1}^{n_1}}E_{(i_k)}^{31}
    \overleftarrow{\prod_{k=2}^{p}}E_{(i_k)}^{32}
    \tilde B_2(\m_1)
     \overrightarrow{\prod_{k=2}^{q+1}}E_{(i_k)}^{23}E_{(j)}^{23}
     \overrightarrow{\prod_{q+2}^{p}}E_{(i_k)}^{23}
     \overrightarrow{\prod_{k=p+1}^{n_1}}E_{(i_k)}^{13}|0\rangle \nonumber\\
&&\quad\times
G^{(0)}(\{\l_k\}_{(p,n_1)},i_2,\ldots,i_{q+1},j,i_{q+2},\ldots,i_{n_1})
   \nonumber\\&
 =&-(2\cosh\eta)^{{(p-1)(p-2)+(n-p)(n-p-1)\over 2}}
 \prod_{l=1}^{p}\prod_{k=p+1}^{n_1}a(\l_l,\x_{i_k})
 \sum^N_{j\ne i_2,\ldots,i_{n_1}}(-1)^{q}\,
 b^+(\m_1,\x_j)
 \nonumber\\ &&\times
 \prod_{l=2}^p a(\m_1,\x_{i_l})
 \prod_{k\ne j,i_{p+1},\ldots,i_{n_1}}^Na^{-1}(\x_k,\x_j)\,
 \nonumber\\ &&\times
 \mbox{det}{\cal B}^-_{p}(\l_1\ldots,\l_p;\x_{i_2},\ldots,\x_{i_{q+1}},
      \x_j,\x_{i_{q+2}},\ldots,\x_{i_{p}})
 \nonumber\\ &&\times
 \prod_{l=p+1}^{n_1} a(\m_1,\x_{i_l})
 \mbox{det}{\cal B}^-_{n-p}(\l_{p+1},\ldots,\l_{n_1};
      \x_{i_{p+1}},\ldots,\x_{i_{n_1}})
      . \label{eq:G-1-1}
\end{eqnarray}
Let $\l_k\, (k=1,\dots,n)$ label the row and $\x_l\,
(l=i_2,\ldots,j,\ldots,i_{p}) $ label the column of the matrix
${\cal B}_{p}$. From (\ref{eq:G-0}), one sees that the column
indices in (\ref{eq:G-1-1}) satisfy the sequence
$i_2<\ldots<j<\ldots<i_{p}$. Therefore, moving the $j$th column in
the matrix ${\cal B}_{p}$ to the first column, we have
\begin{eqnarray}
&& G^{(1)}(\{\l_k\}_{(p,n_1)},\m_1,i_2,\ldots,i_{n_1})\nonumber\\
&&=-(2\cosh\eta)^{{(p-1)(p-2)+(n-p)(n-p-1)\over 2}}
  \prod_{l=1}^{p}\prod_{k=p+1}^{n_1}a(\l_l,\x_{i_k})
 \sum^N_{j\ne i_2,\ldots,i_{n_1}}\,
 b^+(\m_1,\x_j)
 \nonumber\\ &&\quad\times
 \prod_{l=2}^p a(\m_1,\x_{i_l})
 \prod_{k\ne j,i_{p+1},\ldots,i_{n_1}}^Na^{-1}(\x_k,\x_j)\,
 \mbox{det}{\cal B}^-_{p}(\l_1,\ldots,\l_p;\x_j,\x_{i_2},\ldots,\x_{i_{n_1}})
 \nonumber\\ && \quad\times
 \mbox{det}{\cal B}^-_{n-p}(\l_{p+1},\ldots,\l_{n_1};
      \x_{i_{p+1}},\ldots,\x_{i_{n_1}})
 \nonumber\\ &&
 =-(2\cosh\eta)^{{(p-1)(p-2)+(n_1-p)(n_1-p-1)\over 2}}
 \mbox{det}({\cal B}^-)^{(1)}_{p}(\l_1,\ldots,\l_p;\m_1,\x_{i_2},\ldots,\x_{i_{p}})
 \nonumber\\ && \quad\times
 \prod_{l=p+1}^{n_1} a(\m_1,\x_{i_l})
 \mbox{det}{\cal B}^-_{n_1-p}(\l_{p+1},\ldots,\l_{n_1};
      \x_{i_{p+1}},\ldots,\x_{i_{n_1}}),
 \label{eq:G-1}
\end{eqnarray}
where the matrix $({\cal B}^-)^{(1)}_{p}(\{\l_k\},\m_1,
\x_{i_2},\ldots,\x_{i_{p}})$ is given by
\begin{eqnarray}
(({\cal B}^-)^{(1)}_{p})_{\alpha\beta}
 &=&\prod_{k=p+1}^{n_1} a(\l_\a,\x_{i_k})\,a(\m_1,\x_{i_\beta})
 ({\cal B}^-_{p})_{\alpha\beta} \,
 (1\leq\a\leq p\,\mbox{ and }2\leq\beta\leq p), \label{eq:B-1-0}\no\\ \\
(({\cal B}^-)^{(1)}_{p})_{\alpha 1}&=&
\prod_{k=p+1}^{n_1}a(\l_\a,\x_{i_k}) \sum^N_{j\ne
i_2,\ldots,i_{n_1}}
 b^+(\m_1,\x_j)b^-(\l_{\alpha},\x_j)
 \prod_{\gamma=1}^{\alpha-1}a(\l_\gamma,\x_j) \nonumber\\ &&\times
 \prod_{k\ne j,i_{p+1},\ldots,i_{n_1}}^N a^{-1}(\x_{k},\x_j)
 \quad\quad (1\leq\a\leq p). \label{eq:B-1-1}
\end{eqnarray}
Using the properties of determinant, one finds that if
$j=i_2,\ldots,i_{p}$, the corresponding terms in (\ref{eq:B-1-1})
contribute zero to the determinant. Thus, without changing the
determinant of the matrix ${\cal B}^{(1)}_{p}$, the elements
$({\cal B}^{(1)}_{p})_{\alpha 1}$ in (\ref{eq:B-1-1}) may be
replaced by
\begin{eqnarray}
&&(({\cal B}^-)^{(1)}_{p})_{\alpha 1}=e^{\m_1}f(\m_1)
 \equiv e^{\m_1}\cdot
 \prod_{k=p+1}^{n_1}a(\l_\a,\x_{i_k})
 \sum^N_{j\ne i_{p+1},\ldots,i_{n_1}} e^{-\m_1}
 \nonumber\\ &&\times
 b^+(\m_1,\x_{j})b^-(\l_\a,\x_{j}) \prod_{\gamma=1}^{\alpha-1}a(\l_\g,\x_j)
 \prod_{k\ne j,i_{p+1},\ldots,i_{n_1}}^N
  a^{-1}(\x_{k},\x_j).\label{eq:B-1-2}
\end{eqnarray}
Thanks to the Bethe ansatz equation (\ref{eq:BAE}), we may
construct the function
\begin{eqnarray}
&&{\cal M}^\pm_{\alpha\beta}
 =
 {b^\pm(\l_\a,\m_\b)\over a(\l_\a,\m_\b)}
 \prod_{\g=1}^{\a-1}a^{-1}(\l_\g,\m_\b)
\prod_{\epsilon=1}^{\b-1} a^{-1}(\m_\e,\m_\b)
  \nonumber\\&&\quad \quad\times
  \left[\prod_{j=p+1}^{n_1}
  a(\m_\b,\l_j^{(1)})
 -\prod_{k=1}^Na(\m_\beta,\x_k)
  \prod_{j=p+1}^{n_1}a^{-1}(\m_\beta,\x_{i_j})
  a(\l_\a,\x_{i_j})
  \right]\nonumber\\
  &&+\sum_{j=p+1}^{n_1}\left[b^\mp(\m_\b,\l_j^{(1)})
  b^\pm(\l_\a,\l_{j}^{(1)})
  \prod_{\g=1}^{\a-1}a(\l_\g,\l_j^{(1)})
  \right. \nonumber\\ &&\quad \quad\times\left.
  \prod_{\e=1}^{\b-1}a(\m_\e,\l_j^{(1)})
  \prod_{k=p+1,\ne j}^{n_1}a^{-1}(\l_k^{(1)},\l_j^{(1)})\right]
  \nonumber\\&& \mbox{} -
  \sum_{\g=1}^{\a-1}
  {b^\mp(\m_\b,\l_\g)\over a(\m_\b,\l_\g)}
  {b^\pm(\l_\a,\l_\g)\over a(\l_\a,\l_\g)}
 \prod_{\iota=1,\ne \g}^{\a-1}a^{-1}(\l_\g,\l_\iota)
   \prod_{\e=1}^{\b-1}a^{-1}(\l_\g,\m_\e)
 \nonumber\\&&\quad \quad\times
  \left[\prod_{j=p+1}^{n_1}
  a(\l_\g,\l_j^{(1)})
 -\prod_{k=1}^Na(\l_\g,\x_k)
  \prod_{j=p+1}^{n_1}a^{-1}(\l_\g,\x_{i_j})
  a(\l_\a,\x_{i_j})
  \right]
    \nonumber\\&& \mbox{} -
  \sum_{\e=1}^{\b-1}
  {b^\mp(\m_\b,\m_\e)\over a(\m_\b,\m_\e)}
  {b^\pm(\l_\a,\m_\e)\over a(\l_\a,\m_\e)}
 \prod_{\g=1}^{\a-1}a^{-1}(\m_\e,\l_\g)
 \prod_{\varepsilon=1,\ne\e}^{\b-1}a^{-1}(\m_\e,\m_\varepsilon)
 \nonumber\\&&\quad \quad\times
  \left[\prod_{j=p+1}^{n_1}a(\m_\e,\l_j^{(1)})
 -\prod_{k=1}^Na(\m_\e,\x_k)
  \prod_{j=p+1}^{n_1} a^{-1}(\m_\e,\x_{i_j})
  a(\l_\a,\x_{i_j})
  \right],
 \label{eq:M-ab}
\end{eqnarray}
where $\l_j^{(1)}$ ($j=p+1,\ldots,n$) satisfy the NBAE
(\ref{eq:BAE-nested}). By direct computation, one sees that the
residues of $e^{-\m_1}{\cal M}^-_{\a 1}$ at points
$\m_1=\l^{(1)}_j-\eta\,\mbox{mod}(i\pi)$,
$\m_1=\l_\g\,\mbox{mod}(i\pi)\, (\g=1,\ldots,\a-1)$ and
$\m_1=\m_\e\,\mbox{mod}(i\pi)\, (\e=1,\ldots,\b-1)$ are zero.
Moreover the residues of $e^{-\m_1}{\cal M}^-_{\a 1}$ at the
points $\m_1=\l_\a\,\mbox{mod}(i\pi)$ are also zero because
$\l_\a\,\mbox{mod}(i\pi)$ is a solution of the BAE (\ref{eq:BAE}).
Then comparing (\ref{eq:B-1-2}) with (\ref{eq:M-ab}),  one finds
that as functions of $\m_1$, the functions $f(\m_1)$ and
$e^{-\m_1}{\cal M}^-_{\a 1}$ have the same residues at the simple
poles $\m_1=\x_j-\eta\,\mbox{mod}(i\pi)$ $(j\ne
i_{p+1},\ldots,i_{n_1})$, and that when $\m_1\rightarrow \infty$,
they are bounded. Therefore, according to the properties of the
analytic functions (Liouville theorem), we have $(({\cal
B}^-)^{(1)}_{p})_{\alpha 1}= {\cal M}_{\a1}^-$.

Then, by using the function $G^{(0)},G^{(1)}$ and the intermediate
function (\ref{de:Gm-Gm-1}) repeatedly, we obtain $G^{(m)}$
$(m\leq p)$:
\begin{eqnarray}
&&G^{(m)}(\{\l_k\}_{(p,n_1)},\m_1,\ldots,\m_m,i_{m+1},\ldots,i_{n_1})
\nonumber\\
 &=&(-1)^{m} (2\cosh\eta)^{{p(p-1)-m(2p-m-1)+(n_1-p)(n_1-p-1)\over 2}}
 \nonumber\\&&\quad\times
 \mbox{det}({\cal B}^-)^{(m)}_p(\l_1,\ldots,\l_p;\m_1,\ldots,\m_{m},
  i_{m+1},\ldots,i_{p})
 \nonumber\\ && \quad\times
 \prod_{l=1}^{m}\prod_{k=p+1}^{n_1}a(\m_l,\x_{i_k})
 \mbox{det}{\cal B}^-_{n_1-p}(\l_{p+1},\ldots,\l_{n_1};
      \x_{i_{p+1}},\ldots,\x_{i_{n_1}})
 \label{eq:G-m}
\end{eqnarray}
with the matrix elements 
\begin{eqnarray}
(({\cal B}^-)^{(m)}_{p})_{\alpha\beta}&=&\prod_{\e=1}^{\b-1}
 a(\m_\e,\x_{i_\beta})({\cal B}^-_{p})_{\a\b},
 \quad\quad (1\leq \a\leq p,\ m<\beta\leq p),\nonumber\\
(({\cal B}^-)^{(m)}_{p})_{\alpha\beta}
 &=&{\cal M}^-_{\alpha\beta},
 \quad\quad\quad\quad\quad\quad\quad\quad\
 (1\leq \a\leq p,\ 1\leq\beta\leq m).
\end{eqnarray}
(\ref{eq:G-m}) can be proved by induction. Firstly from
(\ref{eq:G-1}), (\ref{eq:B-1-0}) and (\ref{eq:M-ab}),
(\ref{eq:G-m}) is true for $m=1$. Assume (\ref{eq:G-m}) for
$G^{(m-1)}$. Let us show (\ref{eq:G-m}) for general $m$.
Substituting $G^{(m-1)}$ and (\ref{eq:B-expect}) into intermediate
function (\ref{de:Gm-Gm-1}), we have
\begin{eqnarray}
&&G^{(m)}(\{\l_k\}_{(p,n_1)},\m_1,\ldots,\m_{m},i_{m+1},\ldots,i_{n_1})
=-(-1)^{q}(2\cosh\eta)^{-(p-m)}
 \nonumber\\&& \quad\times
 \sum_{j\ne i_{m+1},\ldots,i_{n_1}}^N
 b^+(\m_{m},\x_j)\prod_{l=m+1}^{n_1} a(\m_{m},\x_{i_l})
 \prod_{k\ne j,p+1,\ldots,n}^Na^{-1}(\x_k,\x_j) \nonumber\\
 &&\quad\times
 G^{(m-1)}(\{\l_k\}_{(p,n_1)},\m_1,\ldots,\m_{m-1},i_{m+1},
 \ldots, i_{m+p},j,i_{m+p+1},\ldots,i_{n_1})\nonumber\\
 &&=-(2\cosh\eta)^{-(p-m)}\
 \sum_{j\ne i_{m+1},\ldots,i_{n_1}}^N
 b^+(\m_{m},\x_j)\prod_{l=m+1}^{n_1} a(\m_{m},\x_{i_l})
 \prod_{k\ne j,p+1,\ldots,n_1}^Na^{-1}(\x_k,\x_j)\nonumber\\
 &&\quad\times
 G^{(m-1)}(\{\l_k\}_{(p,n_1)},\m_1,\ldots,\m_{m-1},j,i_{m+1},
 \ldots, i_{n_1})\nonumber\\
 &&=(-1)^{m}(2\cosh\eta)^{{p(p-1)-m(2p-m-1)+(n_1-p)(n_1-p+1)\over 2}}\,
 \nonumber\\ && \quad\times
 \mbox{det}({\cal B}^-)^{(m)}_{p}(\l_1,\ldots,\l_p;
      \m_1\ldots,\m_{m},i_{m+1},\ldots,i_{p})
 \nonumber\\ && \quad\times
 \prod_{l=1}^{m}\prod_{k=p+1}^{n_1}a(\m_l,\x_{i_k})
 \mbox{det}{\cal B}^-_{n_1-p}(\l_{p+1},\ldots,\l_{n_1};
      \x_{i_{p+1}},\ldots,\x_{i_{n_1}}),
 \label{eq:Gm-induction}
\end{eqnarray}
where the matrix elements $({\cal B}^{(m)}_{p})_{\alpha\beta}$ are
given by
\begin{eqnarray}
(({\cal B}^-)^{(m)}_{p})_{\alpha\beta}&=&
 \prod_{\e=1}^{\b-1} a(\m_\e,\x_{i_\beta})
 \prod_{k=p+1}^{n_1} a(\l_\a,\x_{i_k})
 ({\cal B}^-_{p})_{\alpha\beta}
 \quad\quad (1\leq \a\leq p,\, m<\beta\leq p),\nonumber\\
(({\cal B}^-)^{(m)}_{p})_{\alpha\beta}
 &=&{\cal M}^-_{\alpha\beta}
 \quad\quad\quad\quad\quad\quad\quad\quad\quad
 \quad\quad\quad\quad\quad\ \
 (1\leq \a\leq p,\, 1\leq \beta<m),
 \nonumber\\
(({\cal B}^-)^{(m)}_{p})_{\alpha m}
 &=&
 \prod_{k=p+1}^{n_1} a(\l_\a,\x_{i_k})
 \sum_{j\ne i_{m+1},\ldots,i_{n_1}}
 b^+(\m_{m},\x_j)b^-(\l_{\alpha},\x_j)
 \prod_{\g=1}^{\a-1}a(\l_\g,\x_j)
 \nonumber\\ &&\quad\times
 \prod_{\e=1}^{\b-1}a(\m_\e,\x_{j})
 \prod_{k\ne j,i_{p+1},\ldots,i_{n_1}}^N a^{-1}(\x_{k},\x_j)
 \quad\quad(1\leq \a\leq p).
 \label{eq:Bm1}
\end{eqnarray}
By the procedure leading to $(({\cal
B}^-)_{p}^{(1)})_{\alpha\beta}$, we prove $(({\cal
B}^-)^{(m)}_{p})_{\alpha m}={\cal M}^-_{\alpha m}$. Therefore we
have proved that the function (\ref{eq:G-m}) holds for all $m\leq
p$.

When $m=p$, we have,
\begin{eqnarray}
&&G^{(p)}(\{\l_k\}_{(p,n_1)},\m_1,\ldots,\m_{p},i_{p+1},\ldots,i_{n_1}) \nonumber\\
 &&=(-1)^{p} (2\cosh\eta)^{{(n_1-p)(n_1-p-1)\over 2}}
 \mbox{det}{\cal M}^-(\l_1,\ldots,\l_{p};\m_1,\ldots,\m_{p})
 \nonumber\\ && \quad\times
 \prod_{l=1}^p\prod_{k=p+1}^{n_1} a(\m_l,\x_{i_k})\
 \mbox{det}{\cal B}^-_{n_1-p}(\l_{p+1},\ldots,\l_{n_1};
      \x_{i_{p+1}},\ldots,\x_{i_{n_1}}),
 \label{eq:G-p}
\end{eqnarray}
where the matrix elements of ${\cal M}^-$ are given by
(\ref{eq:M-ab}).

For later use, we rewrite the element of the matrix ${\cal
M}^\pm_{\a\b}\,(1\leq\a,\b\leq p)$ in the form
\begin{eqnarray}
 {\cal M}^\pm_{\a\b}&=&{\cal F}^\pm_{\a\b}
 +\sum_{\e=1}^\b\prod_{j=p+1}^{n_1} \left(
  a^{-1}(\m_\e,\x_{i_j})
  a(\l_\a,\x_{i_j})\right)\,({\cal G}^\pm)_{\a\b}^\e
  \nonumber\\ &&\mbox{}
 +\sum_{\g=1}^{\a-1}\prod_{j=p+1}^{n_1} \left(
  a^{-1}(\l_\g,\x_{i_j})
  a(\l_\a,\x_{i_j})\right)\,
 ({\cal H}^\pm)^\g_{\a\b},
\end{eqnarray}
where
\begin{eqnarray}
&&{\cal F}^\pm_{\a\b}={b^\pm(\l_\alpha,\m_\b)\over
a(\l_\alpha,\m_\b)}
  \prod_{\g=1}^{\a-1}a^{-1}(\m_\b,\l_\g)
  \prod_{\e=1}^{\b-1}a^{-1}(\m_\b,\m_\e)
  \prod_{j=p+1}^{n_1}a(\m_\b,\l_{j}^{(1)})
  \nonumber\\ &&\mbox{}+
  \sum_{j=p+1}^{n_1}\left[
  b^\mp(\m_\b,\l_j^{(1)})b^\pm(\l_\a,\l_{j}^{(1)})
  \prod_{\g=1}^{\a-1}a(\l_\g,\l_j^{(1)})
  \prod_{\e=1}^{\b-1}a(\m_\e,\l_j^{(1)})
  \prod_{k\ne j}a^{-1}(\l_k^{(1)},\l_j^{(1)})\right]
  \nonumber\\ &&\mbox{}-\sum_{\g=1}^{\a-1}
  {b^\pm(\l_\alpha,\l_\g)\over a(\l_\alpha,\l_\g)}
  {b^\mp(\m_\b,\l_\g)\over a(\m_\b,\l_\g)}
 \prod_{\iota=1,\ne \g}^{\a-1}a^{-1}(\l_\g,\l_\iota)
 \prod_{\e=1}^{\b-1}a^{-1}(\l_\g,\m_\e)
 \prod_{j=p+1}^{n_1}a(\l_\g,\l_{j}^{(1)})
 \nonumber\\ &&  \mbox{}
 -\sum_{\e=1}^{\b-1}
  {b^\pm(\l_\a,\m_\e)\over a(\l_\alpha,\m_\e)}
  {b^\mp(\m_\b,\m_\e)\over a(\m_\b,\m_\e)}
 \prod_{\g=1}^{\a-1}a^{-1}(\m_\e,\l_\g)
 \prod_{\varepsilon=1,\ne \e}^{\b-1}a^{-1}(\m_\e,\m_\varepsilon)
 \prod_{j=p+1}^{n_1}a(\m_\e,\l_{j}^{(1)}),\nonumber \\
%
%
&& ({\cal G}^\pm)^\e_{\a\b}=\left\{\begin{array}{l} \displaystyle
  {b^\pm(\l_\a,\m_\e)\over a(\l_\alpha,\m_\e)}
 \prod_{\g=1}^{\a-1}a^{-1}(\m_\e,\l_\g)
 \prod_{k=1}^N a(\m_\b,\x_k) \quad \quad (\e=\b) \\ \displaystyle
 {b^\pm(\l_\a,\m_\e)\over a(\l_\alpha,\m_\e)}
  {b^\mp(\m_\b,\m_\e)\over a(\m_\b,\m_\e)}
 \prod_{\g=1}^{\a-1}a^{-1}(\m_\e,\l_\g)
 \prod_{\varepsilon=1,\ne \e}^{\b-1}a^{-1}(\m_\e,\m_\varepsilon)
 \prod_{k=1}^N a(\m_\e,\x_k) \\
 ~~~~~~~~~~\quad (1\leq\e\leq \b-1)
 \end{array}\right. \\
 %
 %
 && ({\cal H}^\pm)^\g_{\a\b}={b^\pm(\l_\a,\l_\g)\over a(\l_\alpha,\l_\g)}
  {b^\mp(\m_\b,\l_\g)\over a(\m_\b,\l_\g)}
 \prod_{\iota=1,\ne \g}^{\alpha-1}a^{-1}(\l_\g,\l_\iota)
 \prod_{\e=1}^{\b-1}a^{-1}(\l_\g,\m_\e)
 \prod_{k=1}^N a(\l_\g,\x_k)
 \no\\ &&
 ~~~~~~~~~~~~~\quad (1\leq\g\leq \a-1).
\end{eqnarray}
After a tedious computation similar to that for the supersymmetric
$t$-$J$ model \cite{zsy0511}, we obtain the determinant of the
matrix $\cal M$
\begin{eqnarray}
 &&\mbox{det}{\cal M}^\pm(\{\l_\a\},\{\m_\b\})\nonumber\\
 &&=\mbox{det} {\cal F}^\pm(\{\l_\a\},\{\m_\b\})
 \nonumber\\&&\mbox{}+
 {\sum_{k,l_k,\varrho^k_{m_k}}}'\,
 \prod_{e=1}^k\prod_{g=p+1}^{n_1}
 a^{-l_e}(\m_{e},\x_{i_g})
 \sum_{\stackrel{j_1,\ldots,j_p=1}{j_1\ne\ldots\ne j_p}}^p
 (-1)^{\tau(j_1j_2\ldots j_p)}
 \nonumber\\ &&\quad\times
  \prod_{f'=1}^p\prod_{f=p+1}^{n_1}
  \prod_{t=1}^k\prod_{t'=1}^{l_t}\left[1+
  \delta_{j_{f'}\,\varrho^t_{t'}}
 \left(a(\l_{f'},\x_{i_f})-1\right)\right]
 \nonumber\\ &&\quad\times
 (A_2^\pm)_{1\,j_1}(A_2^\pm)_{2\,j_2}\ldots (A_2^\pm)_{p\,j_p}
 \nonumber\\[2mm] &&\mbox{} +
{\sum_{k,l_k,\rho^k_{m_k}}}'' \,
 \prod_{e=1}^{k}
 \prod_{g=p+1}^{n_1} a^{-l_e}(\l_{e},\x_{i_g})^{}
 \prod_{t=1}^k\prod_{t'=1}^{l_t}
 \prod_{g'=p+1}^{n_1} a(\l_{\rho^t_{t'}},\x_{i_{g'}})\,
  \mbox{det}\, A^\pm_3(\{\l_\a\},\{\m_\b\})
 \nonumber\\ &&\mbox{}+
 {\sum_{k,l_k,\varrho^k_{m_k}}}'\,
 {\sum_{k',l'_{k'},\rho^{k'}_{m'_{k'}}}}''
 \prod_{e=1}^{k}
 \prod_{g=p+1}^{n_1} a^{-l'_e}(\l_{e},\x_{i_g})
 \prod_{t=1}^{k'}\prod_{t'=1}^{l'_{t}}
 \prod_{g'=p+1}^{n_1} \left(a(\l_{\rho^t_{t'}},\x_{i_{g'}})\right)
 \nonumber\\ &&\quad \times
 \, \sum_{\stackrel{j_1,\ldots,j_p=1}{j_1\ne\ldots\ne j_p}}^p
 (-1)^{\tau(j_1j_2\ldots j_p)}
 \nonumber\\&& \quad\times
  \prod_{t=1}^k\prod_{f=p+1}^{n_1}
 \prod_{\stackrel{s=1}{s\ne\{\rho^1_{m}\},\ldots,
          \{\rho^{k}_{m_{k}}\}}}^p
 \prod_{f'=1}^p\prod_{t'=1}^{l_t}\left[1+
 \delta_{f'\,s}\,\delta_{j_{f'}\,\varrho^t_{t'}}
 \left(a^{-1}(\m_{t},\x_{i_f})-1\right)\right]
 \nonumber\\&& \quad\times
  \prod_{t=1}^k\prod_{f=p+1}^{n_1}
  \prod_{\stackrel{s=1}{s\ne\{\rho^1_{m_1}\},\ldots,
          \{\rho^{k}_{m_{k}}\}}}^p
 \prod_{f'=1}^p \prod_{t'=1}^{l_t}
 \left[1+
 \delta_{f'\,s}\,\delta_{j_{f'}\,\varrho^t_{t'}}
 \left(a(\l_{s},\x_{i_f})-1\right)\right]
 \nonumber\\&& \quad\times
 (A_4^\pm)_{1\,j_1}(A_4^\pm)_{2\,j_2}\ldots (A_4^\pm)_{p\,j_p} \label{eq:M-t2}\\
 &\equiv& {\cal T}^\pm_1+{\cal T}^\pm_2+
  {\cal T}^\pm_3+{\cal T}^\pm_4,\label{eq:M-t1t2}
\end{eqnarray}
where $\tau(x_1,\ldots,x_p)=\tau(\s)$, ($\s\in{\cal S}_p$ and
$(x_1,\ldots,x_p)=\s(1,\ldots,p)$), $\tau(\s)=0$ if $\s$ is even
and $\tau(\s)=1$ if $\s$ is odd,
\begin{eqnarray*}
{\sum_{k,l_k,\varrho^k_{m_k}}}'&=& \sum_{k=1}^{p}\left\{
 \sum_{l_k=1}^{p-k+1}{1\over l_k!}\prod_{m_k=1}^{l_k}
 \sum_{\stackrel{\varrho^k_{m_k}=k}
       {\varrho^k_{m_k}\ne\varrho^k_{m_k-1},\ldots,\varrho^k_{1} }}^{p}
 \right\}
 \left\{
 \prod_{r=1}^{k-1}\,\,
 \sum_{l_{k-r}=0}^{p-k+r+1-\sum_{j=0}^{r-1}l_{k-j}}
 {1\over l_{k-r}!}
 \right.\nonumber\\ &&\quad\times \left.
 \prod_{m_{k-r}=1}^{l_{k-r}}
 \sum_{\stackrel{\varrho_{m_{k-r}}^{k-r}=k-r}
       {\stackrel{\varrho_{m_{k-r}}^{k-r}\ne\varrho_{m_{k-r}-1}^{k-r},
       \ldots,\varrho_{1}^{k-r}}
        {\varrho_{m_{k-r}}^{k-r}\ne\{\varrho_{\kappa}^{k-r+1}\},
        \{\varrho_{\kappa}^{k-r+2}\},\ldots,\{\varrho_{\kappa}^{k}\}}}}
       ^{p} \right\},\\
{\sum_{k,l_k,\rho^k_{m_k}}}''&=&
  \sum_{k=1}^{p-1}\left\{
 \sum_{l_k=1}^{p-k}{1\over l_k!}\prod_{m_k=1}^{l_k}
 \sum_{\stackrel{\rho^k_{m_k}=k+1}
       {\rho^k_{m_k}\ne\rho^k_{m_k-1},\ldots,\rho^k_{1} }}^{p}
  \right\}
 \left\{
 \prod_{r=1}^{k-1}\,
 \sum_{l_{k-r}=0}^{p-k+r-\sum_{j=0}^{r-1}l_{k-j}}
 {1\over l_{k-r}!}
 \right. \nonumber\\ &&\quad\times \left.
 \prod_{m_{k-r}=1}^{l_{k-r}}
 \sum_{\stackrel{\rho_{m_{k-r}}^{k-r}=k-r+1}
       {\stackrel{\rho_{m_{k-r}}^{k-r}\ne\rho_{m_{k-r}-1}^{k-r},
       \ldots,\rho_{1}^{k-r}}
        {\rho_{m_{k-r}}^{k-r}\ne\{\rho_{\kappa}^{k-r+1}\},
        \{\rho_{\kappa}^{k-r+2}\},\ldots,\{\rho_{\kappa}^{k}\}}}}
       ^{p}
 \right\},
\end{eqnarray*}
and the elements $(A_i^\pm)_{\a\b}$, $i=2,3,4$, are given by
\begin{eqnarray}
&&(A_2^\pm)_{\a\b}=\left\{\begin{array}{cl}
 {\cal F}^\pm_{\a\b}\quad\quad & \a=1,\dots,p,\,\,
 \b=1,\dots,p,\, \b \ne
 \{\varrho_{m_1}^1\},\ldots, \{\varrho_{m_{n_1}}^{n_1}\}\\[2mm]
 ({\cal G}^\pm)^k_{\a\b}\quad\quad &
  \a=1,\dots,p, \b=\{\varrho^k_{m_k}\} (k=1,2,\ldots,p)
 \end{array},
 \right.
\nonumber\\
\end{eqnarray}
\begin{eqnarray}
&&(A_3^\pm)_{\a\b}=\left\{\begin{array}{cl}
 {\cal F}^\pm_{\a\b}\; & \a=1,\dots,p,\,
 \a\ne \{\rho_{m_1}^1\},\ldots, \{\rho_{m_{n_1-1}}^{n_1-1}\},\,\,
 \b=1,\dots,p\\[2mm]
 \displaystyle
 {1\over\b+1}({\cal H}^\pm)^k_{\a\b}\; &
  \a= \{\rho^k_{m_k}\} (k=1,2,\ldots,p-1),\,\,\b=1,\dots,p
 \end{array},
 \right.
\nonumber\\
\end{eqnarray}
and
\begin{eqnarray}
&&(A_4^\pm)_{\a\b}=\left\{\begin{array}{cl}
 {\cal F}^\pm_{\a\b}\quad & \a=1,\dots,p,\, \a \ne
 \{\rho_{m_1}^1\},\ldots, \{\rho_{m_{n_1-1}}^{n_1-1}\},\\
 & \b=1,\dots,p,\,
     \b\ne\{\varrho_{m_1}^1\},\ldots, \{\varrho_{m_{n_1}}^{n_1}\}\\[2mm]
 ({\cal G}^\pm)^k_{\a\b}\quad &
  \a=1,\dots,p,\,
  \a\ne\{\rho_{m_1}^1\},\ldots, \{\rho_{m_{n_1-1}}^{n_1-1}\},\\ &
  \b=\{\varrho_{m_k}^k\}\,(k=1,2,\ldots,n)\\[2mm]
  \displaystyle
 {1\over\b+1}({\cal H}^\pm)^k_{\a\b}\quad &
 \a=\{\rho_{m_k}^k\}\, (k=1,\ldots n_1-1),\,\,
 \b=1,\dots,p
 \end{array}
 \right.,\nonumber\\
\end{eqnarray}
respectively.

Thus by using (\ref{eq:M-t1t2}), the function $G^{(p)}$
(\ref{eq:G-p}) becomes
\begin{eqnarray}
 && G^{(p)}(\{\l_k\}_{(p,n_1)},\m_1,\ldots,\m_p,i_{p+1},\ldots,i_{n_1})
 \nonumber\\ &&=
 (-1)^{p}(2\cosh\eta)^{{(n_1-p)(n_1-p-1)\over 2}}
 \sum_{j=1}^4{\cal T}^-_j
  \nonumber\\ &&  \quad\times
  \prod_{l=1}^p\prod_{k=p+1}^{n_1} a(\m_l,\x_{i_k})
  \mbox{det}{\cal B}^-_{n_1-p}(\l_{p+1},\ldots,\l_{n_1};
      \x_{i_{p+1}},\ldots,\x_{i_{n_1}}) 
  \nonumber\\
 &&\equiv \sum_{j=1}^4
 (G^-)^{(p)}_j(\{\l_k\}_{(p,n_1)},\m_1,\ldots,\m_p,i_{p+1},\ldots,i_{n_1}).
 \label{eq:Gp-G1G2}
\end{eqnarray}

\vskip12pt
\begin{itemize}
\item $m\geq p+1$
\end{itemize}

Then we compute the intermediate functions $G^{(m)}$ for $m\geq
p+1$. Similar to the $m\leq p$ case, inserting a complete set and
noticing (\ref{eq:Gp-G1G2}), we have
\begin{eqnarray}
&&G^{(m)}(\{\l_k\}_{(p,n_1)},\m_1,\ldots,\m_m,i_{m+1},\ldots,i_{n_1})\nonumber\\
 &&=\sum_{j\ne i_{m+1},\ldots,i_{n_1}}^N
     \langle0|\overleftarrow{\prod_{k=m+1}^{n_1}}E_{(i_k)}^{31}
    \tilde B_1(\m_m)
     \overrightarrow{\prod_{k=m+1}^{m+q}}E_{(i_k)}^{13}E_{(j)}^{13}
     \overrightarrow{\prod_{m+q+1}^{n_1}}E_{(i_k)}^{13}|0\rangle
     \nonumber\\
 &&\quad\times
 (G^-)^{(m-1)}(\{\l_k\}_{(p,n_1)},\m_1,\ldots,\m_{m-1},i_{m+1},
 \ldots,i_{m+q},j,i_{m+q+1}\ldots,i_{n_1})
 \nonumber\\
 &&=\sum_{j=1}^4(G^-)^{(m)}_j(\{\l_k\}_{(p,n_1)},\m_1,\ldots,\m_m,i_{m+1},\ldots,i_{n_1}),
 \label{de:Gm-Gm-1234}
\end{eqnarray}
where $(G^-)^{(m)}_j$'s correspond to $(G^-)^{(p)}_j$'s in
(\ref{eq:Gp-G1G2}), respectively.

We first compute $(G^-)^{(m)}_1$. With the help of the expression
of $\tilde B_1$ (\ref{eq:B1-tilde}), we have
\begin{eqnarray}
&&\langle0|\overleftarrow{\prod_{k=m+1}^{n_1}}E_{(i_k)}^{31}
    \tilde B_1(\m_m)
     \overrightarrow{\prod_{k=m+1}^{m+q}}E_{(i_k)}^{13}E_{(j)}^{13}
     \overrightarrow{\prod_{m+q+1}^{n_1}}E_{(i_k)}^{13}|0\rangle
\nonumber\\
 &=&-(-1)^{q}(2\cosh\eta)^{-(n_1-m)}\,
 b^+(\m_m,\x_j)\prod_{l=m+1}^{n_1} a(\m_m,\x_{i_l})
 \prod_{k\ne j}^Na^{-1}(\x_k,\x_j)
 . \label{eq:B1-expect}
\end{eqnarray}
When $m=p+1$,  by using the expressions (\ref{eq:Gp-G1G2}) and
(\ref{eq:B1-expect}), the intermediate function $(G^-)^{(p+1)}_1$
is given by

\begin{eqnarray}
&&(G^-)^{(p+1)}_1(\{\l_k\}_{(p,n_1)},\m_1,\ldots,\m_{p+1},i_{p+2},\ldots,i_{n_1})
 \nonumber\\ &&=
 \sum_{j\ne i_{p+2},\ldots,i_{n_1}}^N
     \langle0|\overleftarrow{\prod_{k=p+2}^{n_1}}E_{(i_k)}^{31}
    \tilde B_1(\m_{p+1})
     \overrightarrow{\prod_{k=p+2}^{p+q+1}}E_{(i_k)}^{13}E_{(j)}^{13}
     \overrightarrow{\prod_{p+q+2}^{n_1}}E_{(i_k)}^{13}|0\rangle
     \nonumber\\
 &&\quad\times
 (G^-)^{(p)}_1(\{\l_k\}_{(p,n_1)},\m_1,\ldots,\m_{p},i_{p+2},
 \ldots,i_{p+q+1},j,i_{p+q+2}\ldots,i_{n_1})
 \nonumber\\
 &&=(-1)^{m} (2\cosh\eta)^{{(n_1-p)(n_1-p-1)-2(n_1-p-1)\over 2}}
  \mbox{det}{\cal F}(\l_1,\ldots,\l_p;\m_1,\ldots,\m_p)
 \nonumber\\ &&\quad \times
 \mbox{det}({\cal B}^-)^{(p+1)}_{n_1-p}(\l_{p+1},\ldots,\l_{n_1};
      \m_{p+1};\x_{i_{p+2}},\ldots,\x_{i_{n_1}}),
 \label{eq:G-m-n-p}
\end{eqnarray}
where the matrix elements $(({\cal
B}^-)^{(m)}_{n_1-p})_{\alpha\beta}$ $(p+1\leq\a,\b\leq n_1)$
\begin{eqnarray}
\left(({\cal B}^-)^{(p+1)}_{n_1-p}\rt)_{\alpha\beta}&=&
 \prod_{\e=1}^{\b-1}
 a(\m_{\e},\x_{i_\beta})({\cal B}^-_{n_1-p})_{\a\b},
 \quad\quad \mbox{for } p+1<\beta\leq n_1,\nonumber\\
 \lt(({\cal B}^-)^{(p+1)}_{n_1-p}\rt)_{\a\,p+1}&=&
  \sum^N_{j\ne i_{p+2},\ldots,i_{n_1}}
 b^+(\m_{p+1},\x_j)b^-(\l_{\a},\x_j)
 \prod_{\g=p+1}^{\a-1}a(\l_\g,\x_j)
 \nonumber\\ &&\times
 \prod_{\e=1}^{p}a(\m_\e,\x_j)
 \prod_{k\ne j}^N a^{-1}(\x_{k},\x_j).
 \label{eq:B-p+1}
\end{eqnarray}
Then by using similar procedure as the $m\leq p$ case, one prove
that $\lt(({\cal B}^\pm)^{(p+1)}_{n_1-p}\rt)_{\a\,p+1}$ is equal
to the function $({\cal N}_1)_{\a\b}$ ($p+1\leq\a,\b\leq n_1$)
\begin{eqnarray}
({\cal N}^\pm_1)_{\a\b}&=&
 {b^\pm(\l_\a,\m_\b)\over a(\l_\a,\m_\b)}
 \prod_{\g=p+1}^{\a-1} a^{-1}(\m_\b,\l_\g)
 \prod_{\e=1}^{\b-1}a^{-1}(\m_\b,\m_\e)
  \nonumber\\&&\quad \times
  \left[\prod_{j=p+1}^{n_1} a(\m_\b,\l_j^{(1)})
 -\prod_{k=1}^Na(\m_\b,\x_k)
  \right]\nonumber\\
  &&+\sum_{j=p+1}^{n_1}\left[b^\mp(\m_\b,\l_j^{(1)})
  b^\pm(\l_\a,\l_{j}^{(1)})
  \prod_{\g=p+1}^{\a-1}a(\l_\g,\l_j^{(1)})
 \right. \nonumber\\ &&\quad \times\left.
  \prod_{\e=1}^{\b-1}a(\m_\e,\l_j^{(1)})
  \prod_{k\ne j=p+1}^{n_1}a^{-1}(\l_k^{(1)},\l_j^{(1)})\right]
    \nonumber\\&& \mbox{} -
  \sum_{\e=1}^{\b-1}{b^\pm(\l_\a,\m_\e)\over a(\l_\a,\m_\e)}
 \prod_{\g=p+1}^{\a-1}a^{-1}(\m_\e,\l_\g)
   \prod_{\varepsilon=1,\ne\e}^{\b-1}a^{-1}(\m_\e,\m_\varepsilon)
 \nonumber\\&&\quad \quad\times
  \left[\prod_{j=p+1}^{n_1}a(\m_\e,\l_j^{(1)})
 -\prod_{k=1}^N a(\m_\e,\x_k)\right].
 \label{eq:M-ab-n}
\end{eqnarray}
Moreover, with a similar procedure, one may prove that for any
$p+1\leq m\leq n_1$, the function $(G^-)^{(m)}_1$ can be written
as
\begin{eqnarray}
&&(G^-)^{(m)}_1(\{\l_k\}_{(p,n_1)},\m_1,\ldots,\m_{m},i_{m+1},\ldots,i_{n_1}) \nonumber\\
 &&=(-1)^{m} 2^{{(n_1-p)(n_1-p-1)-(m-p)(2n_1-m-p-1)\over 2}}
 \mbox{det}{\cal F}^-(\l_1,\ldots,\l_{p};\m_1,\ldots,\m_{p})
 \nonumber\\ && \quad\times
 \mbox{det}({\cal B}^-)^{(m)}_{n_1-p}(\l_{p+1},\ldots,\l_{n_1};
      \x_{i_{p+1}},\ldots,\x_{i_{n_1}}),
 \label{eq:G-m-n-p}
\end{eqnarray}
where the matrix elements $(({\cal
B}^-)^{(m)}_{n_1-p})_{\alpha\beta}$ $(p+1\leq\a,\b\leq n_1)$
\begin{eqnarray}
(({\cal B}^-)^{(m)}_{n_1-p})_{\alpha\beta}&=&\prod_{\e=1}^{\b-1}
 a(\m_\e,\x_{i_\beta})({\cal B}^-_{p})_{\a\b},
 \quad\quad \mbox{for } m<\beta\leq n_1,\nonumber\\
(({\cal B}^-)^{(m)}_{n_1-p})_{\alpha\beta}&=&({\cal
N}_1^-)_{\alpha\beta},
 \quad\quad\quad\quad\quad\quad\,\,\quad \mbox{for } p+1\leq\beta\leq m.
\end{eqnarray}
Therefore when $m=n_1$, we obtain
\begin{eqnarray}
 && (G^-)_1^{(n_1)}(\{\l_j\}_{(p,n_1)},\{\m_k\}_{(p,n_1)})
 =(-1)^{n_1}\,\,
 \mbox{det}{\cal F}^-(\l_1,\ldots,\l_{p};\m_1,\ldots,\m_{p})
 \nonumber\\ && \times
 \mbox{det}\,{\cal N}_1^-(\l_{p+1},\ldots,\l_{n_1};
      \m_{{p+1}},\ldots,\m_{{n_1}}).
 \label{eq:results-G1}
\end{eqnarray}

Similarly, the function $(G^-)_2^{(n_1)}$ is given by
\begin{eqnarray}
 &&(G^-)_2^{(n_1)}(\{\l_j\}_{(p,n_1)},\{\m_k\}_{(p,n_1)})=(-1)^{n_1}
{\sum_{k,l_k,\varrho^k_{m_k}}}'
 \sum_{\stackrel{j_1,\ldots,j_p=1}{j_1\ne\ldots\ne j_p}}^p
 (-1)^{\tau(j_1j_2\ldots j_p)}
 \nonumber\\ &&\quad\times
 (A_2^-)_{1\,j_1}(A_2^-)_{2\,j_2}\ldots (A_2^-)_{p\,j_p}
 \mbox{det}\,{\cal N}^-_2(\l_{p+1},\ldots,\l_{n_1};
      \m_{{p+1}},\ldots,\m_{{n_1}};\{l_k\}) \no\\
 \label{eq:results-G2}
\end{eqnarray}
with
\begin{eqnarray}
&&({\cal N}^\pm_2)_{\a\b}=
 {b^\pm(\l_\a,\m_\b)\over a(\l_\a,\m_\b)}
 \prod_{e=1}^k a^{l_e-1}(\m_\b,\m_e)
 \prod_{\e=k+1}^{\b-1}a^{-1}(\m_\b,\m_\e)
 \prod_{\g=p+1}^{\a-1}a^{-1}(\m_\b,\l_\g)
 \nonumber\\ &&\,\times
  \prod_{f'=1}^p
 \prod_{t=1}^{k}\prod_{t'=1}^{l_t}
 \left[1+\delta_{j_{f'}\,\varrho^t_{t'}}
 \left(a^{-1}(\m_\b,\l_{f'})-1\right)\right]
  \left[\prod_{j=p+1}^{n_1}
  a(\m_\b,\l_j^{(1)})
 -\prod_{l=1}^N a(\m_\b,\x_l)\right]
 \nonumber\\ &&\mbox{}+
 \sum_{\theta=p+1}^{n_1}b^\mp(\m_\b,\l_\theta^{(1)})
  b^\pm(\l_\a,\l_\theta^{(1)})
 \prod_{e=1}^ka^{l_e-1}(\m_e,\l_\theta^{(1)})
 \prod_{\e=k+1}^{\b-1}a(\m_\e,\l_\theta^{(1)})
 \nonumber\\ &&\quad\times
 \prod_{\g=p+1}^{\a-1}a(\l_\g,\l_\theta^{(1)})
   \prod_{\vartheta=p+1,\ne\theta}^{n_1}
  a^{-1}(\l_\vartheta,\l_\theta^{(1)})
 \nonumber\\ &&\quad\times
  \prod_{f'=1}^p
 \prod_{t=1}^{k}\prod_{t'=1}^{l_t}
 \left[1+\delta_{j_{f'}\,\varrho^t_{t'}}
 \left(a(\l_{f'},\l_\theta^{(1)})-1\right)\right]
 \nonumber\\ &&\mbox{}-
 \sum_{\e=k+1}^{\b-1}
 {b^\mp(\m_\b,\m_\e)\over a(\m_\b,\m_\e)}
 {b^\pm(\l_\a,\m_\e)\over a(\l_\a,\m_\e)}
 \prod_{e=1}^k a^{l_e-1}(\m_\e,\m_e)
 \prod_{\varepsilon=k+1,\ne \e}^{\b-1}
  a^{-1}(\m_\e,\m_\varepsilon)
 \nonumber\\ &&\quad\times
 \prod_{\g=p+1}^{\a-1}a^{-1}(\m_\e,\l_\g)
  \left[\prod_{j=p+1}^{n_1}
  {a(\m_\e,\l_j^{(1)})}-\prod_{l=1}^N{a(\m_\e,\x_l)}\right]
 \nonumber\\ &&\quad\times
  \prod_{f'=1}^p
 \prod_{t=1}^{k}\prod_{t'=1}^{l_t}
 \left[1+\delta_{j_{f'}\,\varrho^t_{t'}}
 \left({a^{-1}(\m_\e,\l_{f'})}-1\right)\right]
 \nonumber\\ &&\mbox{}+
\sum_{e=1}^{k}g^\pm_2(\m_\b,l_e),
\end{eqnarray}
where the function $g^\pm_2(\m_\b,l_e)=0 $ when $l_e=1$; when
$l_e$=0,
\begin{eqnarray}
&&g_2^\pm(\m_\b,l_e)= -{b^\mp(\m_\b,\m_e)\over a(\m_\b,\m_e)}
{b^{\pm}(\l_\a,\m_e)\over a(\l_\a,\m_e)}
 \prod_{e'=1,\ne e}^k a^{l_{e'}-1}(\m_{e}\m_{e'})
 \prod_{\e=k+1}^{\b-1}a^{-1}(\m_e-\m_\e)
 \nonumber\\ &&\quad\times
 \prod_{\g=p+1}^{\a-1}{a^{-1}(\m_e,\l_\g)}
  \left[\prod_{j=p+1}^{n_1}
  {a(\m_e,\l_j^{(1)})}
 -\prod_{l=1}^N a(\m_e,\x_l)\right]
 \nonumber\\ &&\quad\times
  \prod_{f'=1}^p
 \prod_{t=1}^{k}\prod_{t'=1}^{l_t}
 \left[1+\delta_{j_{f'}\,\varrho^t_{t'}}
 \left(a(\m_e,\l_{f'})-1\right)\right];
\end{eqnarray}
when $l_e\geq 2$,
\begin{eqnarray}
&& g_2(\m_\b,l_e)= -\sum_{k=0}^{l_e-2}{1\over k!}
 {e^{\mp\m_\b}\over \sinh(\m_\b-\m_e+\eta)^{l_e-k-1}}
 {d^{k}\over d\chi^{k}_2}\left\{
 \sinh^{l_e-1}(\m_\b-\m_e)\right.
 \nonumber\\ &&\quad\times
 {e^{\pm\m_\b}b^\pm(\l_\a,\m_\b)\eta\over a(\l_\a,\m_\b)}
 \prod_{e'=1,\ne e}^k a^{l_{e'}-1}(\m_\b,\m_{e'})
 \prod_{\e=k+1}^{\b-1}{a^{-1}(\m_\b,\m_\e) }
 \prod_{\g=p+1}^{\a-1}{a^{-1}(\m_\b,\l_\g)}
 \nonumber\\ &&\quad\times
  \prod_{f'=1}^p
 \prod_{t=1}^{k}\prod_{t'=1}^{l_t}
 \left[1+\delta_{j_{f'}\,\varrho^t_{t'}}
 \left({a^{-1}(\m_\b,\l_{f'})}-1\right)\right]
 \nonumber\\ &&\quad\times \left.
  \left[\prod_{j=p+1}^{n_1}
  {a(\m_\b,\l_j^{(1)})}
 -\prod_{l=1}^N{a(\m_\b,\x_l)}\right],
  \right\}_{\chi_2=0}
\end{eqnarray}
and $\chi_2\equiv \chi_2(\m_\b)=\sinh(\m_\b-\m_e+\eta)$.

\vskip12pt

The function $(G^-)_3^{(n_1)}$ is given by
\begin{eqnarray}
&&(G^-)_3^{(n_1)}(\{\l_j\}_{(p,n_1)},\{\m_k\}_{(p,n_1)}) \no\\&=&
(-1)^{n_1}{\sum_{k,l_k,\rho^k_{m_k}}}''
 \mbox{det}\, A^-_3(\l_{\s(1)},\ldots,\l_{\s(p)};
         \m_{\s'(1)},\ldots,\m_{\s'(p)})
 \nonumber\\ && \times
 \mbox{det}\,{\cal N}^-_3(\l_{p+1},\ldots,\l_{n_1};
      \m_{{p+1}},\ldots,\m_{{n_1}};\{l_k\})
 \label{eq:results-G3}
\end{eqnarray}
with
\begin{eqnarray}
 &&({\cal N}^\pm_3)_{\a\b}=
 {b^\pm(\l_\a,\m_\b)\over a(\l_\a,\m_\b)}
 \prod_{e=1}^k a^{l_e}(\m_\b,\l_e)
 \prod_{t=1}^k\prod_{t'=1}^{l_t}
 {a^{-1}(\m_\b,\l_{\rho^t_{t'}}) }
 \prod_{\g=p+1}^{\a-1}{a^{-1}(\m_\b,\l_\g)}
 \nonumber\\ &&\quad\times
 \prod_{\e=1}^{\b-1}{a^{-1}(\m_\b,\m_\e)}
  \left[\prod_{j=p+1}^{n_1}
  {a(\m_\b,\l_j^{(1)})}
 -\prod_{l=1}^N{a(\m_\b,\x_l)}\right]
 \nonumber\\ &&\mbox{}+
 \sum_{\theta=p+1}^{n_1}
 {b^\mp(\m_\b,\l_\theta^{(1)})}
 {b^\pm(\l_\a,\l_\theta^{(1)})}
 \prod_{\e=1}^{\b-1}{a(\m_\e,\l_\theta^{(1)})}
 \prod_{\g=p+1}^{\a-1}{a(\l_\g,\l_\theta^{(1)}) }
 \nonumber\\ &&\quad\times
 \prod_{e=1}^k {a^{-l_e}(\m_e,\l_\theta^{(1)})}
  \prod_{t=1}^k\prod_{t'=1}^{l_t}
 {a(\l_{\rho^t_{t'}},\l_\theta^{(1)})}
   \prod_{\vartheta=p+1,\ne\theta}^{n_1}
  {a^{-1}(\l_\vartheta,\l_\theta^{(1)})}
 \nonumber\\ &&\mbox{}-
 \sum_{\e=k+1}^{\b-1}
 {b^\mp(\m_\b,\m_\e)}  {b^\pm(\l_\a,\m_\e)}
 \prod_{e=1}^k{a^{l_e}(\m_\e,\l_e)}
 \prod_{t=1}^k\prod_{t'=1}^{l_t}
 {a^{-1}(\l_{\m_\e,\rho^t_{t'}})}
 \nonumber\\ &&\quad\times
  \prod_{\varepsilon=1,\ne \e}^{\b-1}
  {a^{-1}(\m_\e,\m_\varepsilon)}
 \prod_{\g=p+1}^{\a-1}{a^{-1}(\m_\e,\l_\g)}
  \left[\prod_{j=p+1}^{n_1}
  {a(\m_\e,\l_j^{(1)})}
 -\prod_{l=1}^N{a(\m_\e,\x_l)}\right]
 \nonumber\\ &&\mbox{}+
\sum_{e=1}^{k}g^\pm_3(\m_\b,l_e),
\end{eqnarray}
where the function $g^\pm_3(\m_\b,l_e)$ is given as follows. i.)
when $\prod_{t=1}^k\prod_{t'=1}^{l_t}\delta_{e\, \rho^t_{t'}}=0$,
\begin{eqnarray}
 &&g_3(\m_\b,l_e)=
 -\sum_{k=0}^{l_e-1}{1\over k!}
 {e^{\mp\m_\b}\over \sinh(\m_\b-\l_e+\eta)^{l_e-k}}
 {d^{k}\over d\chi^{k}_3}\left\{\sinh^{l_e}(\m_\b-\l_e)
 {e^{\pm\m_\b}b^\pm(\l_\a,\m_\b)\over a(\l_\a,\m_\b)}
 \right. \nonumber\\ && \quad \times
 \prod_{e'=1,\ne e}^k{a^{l_e}(\m_\b,\l_e)}
 \prod_{t=1}^k\prod_{t'=1}^{l_t}
 {a^{-1}(\m_\b,\l_{\rho^t_{t'}})}
 \prod_{\g=p+1}^{\a-1}{a^{-1}(\m_\b,\l_\g)}
 \nonumber\\ &&\quad\times \left.
 \prod_{\e=1}^{\b-1}{a^{-1}(\m_\b,\m_\e)}
  \left[\prod_{j=p+1}^{n_1} {a(\m_\b,\l_j^{(1)})}
 -\prod_{l=1}^N{a(\m_\b,\x_l)}\right]
 \right\}_{\chi_3=0},
\end{eqnarray}
and $\chi_3\equiv \chi_3(\m_\b)=\sinh(\m_\b-\l_e+\eta)$,
 ii.) when $\prod_{t=1}^k\prod_{t'=1}^{l_t}\delta_{e\,
\rho^t_{t'}}=1$ and $l_e=1$, $g^\pm_3(\m_\b,l_e)=0 $ and  iii.)
when there is an index $\hat t$ ($\hat t\in \{1,\ldots k\}$) and
$\hat t'$ ($\hat t'\in \{1,\ldots l_{\hat t}\}$) such that
$\rho^{\hat t}_{\hat t'}=e$, and $l_e\geq 2$,
\begin{eqnarray}
 &&g_3(\m_\b,l_e)=
 -\sum_{k=0}^{l_e-2}{1\over k!}
 {e^{\mp\m_\b}\over
   \sinh(\m_\b-\l_e+\eta)^{l_e-k-1}}
 {d^{k}\over d\chi^{k}_3}\left\{\sinh^{l_e-1}(\m_\b-\l_e)
 \right. \nonumber\\ && \quad \times
 {e^{\pm\m_\b}b^{\pm}(\l_\a,\m_\b)\over a(\l_\a,\m_\b)}
 \prod_{e'=1,\ne e}^k{a^{l_e}(\m_\b,\l_e)}
 \prod_{t=1,\ne \hat t}^k\, \prod_{t'=1,\ne \hat t'}^{l_t}
 {a^{-1}(\m_\b,\l_{\rho^t_{t'}})}
 \nonumber\\ &&\quad\times \left.
  \prod_{\g=p+1}^{\a-1}{a^{-1}(\m_\b,\l_\g)}
 \prod_{\e=1}^{\b-1}{a^{-1}(\m_\b,\m_\e)}
  \left[\prod_{j=p+1}^{n_1}
  {a(\m_\b,\l_j^{(1)})}
 -\prod_{l=1}^N{a(\m_\b,\x_l)}\right]
 \right\}_{\chi_3=0}. \no\\
\end{eqnarray}

The function $(G^-)^{(n_1)}_4$ is given by
\begin{eqnarray}
 &&(G^-)_4^{(n_1)}(\{\l_j\}_{(p,n_1)},\{\m_k\}_{(p,n_1)})\nonumber\\
 &=&(-1)^{n_1}
 {\sum_{k,l_k,\varrho^k_{m_k}}}'
 {\sum_{k',l'_{k'},\rho^{k'}_{m'_{k'}}}}''
 \sum_{\stackrel{j_1,\ldots,j_p=1}{j_1\ne\ldots\ne j_p}}^p
 (-1)^{\tau(j_1j_2\ldots j_p)}
 \nonumber\\ &&\times
 (A_4^-)_{1\,j_1}(A_4^-)_{2\,j_2}\ldots (A_4^-)_{p\,j_p}
 \nonumber\\ && \times
 \mbox{det}\,{\cal N}^-_4(\l_{p+1},\ldots,\l_{n_1};
      \m_{{p+1}},\ldots,\m_{{n_1}};\{l_k\};\{l'_{k'}\})
 \label{eq:results-G4}
\end{eqnarray}
with
\begin{eqnarray}
&&({\cal N}^\pm_4)_{\a\b}=
 {b^{\pm}(\l_\a,\m_\b)\over a(\l_\a,\m_\b)}
 \prod_{e=1}^k{a^{l'_e}(\m_\b,\l_e)}
 \prod_{t=1}^{k'}\prod_{t'=1}^{l'_t}
 {a^{-1}(\m_\b,\l_{\rho^t_{t'}})}
 \prod_{\g=p+1}^{\a-1}{a^{-1}(\m_\b,\l_\g)}
 \nonumber\\ &&\quad\times
 \prod_{\e=1}^{\b-1}{a^{-1}(\m_\b,\m_\e)}
  \left[\prod_{j=p+1}^{n_1}
  {a(\m_\b,\l_j^{(1)})}
 -\prod_{l=1}^N{a(\m_\beta,\x_l)}\right]
 \nonumber\\ &&\quad\times
  \prod_{t=1}^k
 \prod_{\stackrel{s=1}{s\ne\{\rho^1_{m_1}\},\ldots,
          \{\rho^{k}_{m_{k}}\}}}^p
 \prod_{f'=1}^p\prod_{t'=1}^{l_t}
 \left[1+
 \delta_{f'\,s}\,\delta_{j_{f'}\,\varrho^t_{t'}}
 \left(a(\m_\b,\m_{t})
 -1\right)\right]
 \nonumber\\&& \quad\times
  \prod_{t=1}^k
 \prod_{\stackrel{s=1}{s\ne\{\rho^1_{m_1}\},\ldots,
          \{\rho^{k}_{m_{k}}\}}}^p
 \prod_{f'=1}^p \prod_{t'=1}^{l_t}
 \left[1+
 \delta_{f'\,s}\,\delta_{j_{f'}\,\varrho^t_{t'}}
 \left({a^{-1}(\m_\b,\l_{s})}
 -1\right)\right]
 \nonumber\\ &&\mbox{}+
 \sum_{\theta=p+1}^{n_1}
 {b^\mp(\m_\b,\l_\theta^{(1)})}
 {b^\pm(\l_\a,\l_\theta^{(1)})}
 \prod_{\e=1}^{\b-1}a(\m_\e,\l_\theta^{(1)})
 \prod_{\g=p+1}^{\a-1}a(\l_\g,\l_\theta^{(1)})
 \nonumber\\ &&\quad\times
 \prod_{e=1}^k a^{-l'_e}(\m_e,\l_\theta^{(1)})
  \prod_{t=1}^{k'}\prod_{t'=1}^{l'_t}
 {a(\l_{\rho^t_{t'}},\l_\theta^{(1)})}
   \prod_{\vartheta=p+1,\ne\theta}^{n_1}
  {a^{-1}(\l_\vartheta,\l_\theta^{(1)}) }
   \nonumber\\ &&\quad\times
  \prod_{t=1}^k
 \prod_{\stackrel{s=1}{s\ne\{\rho^1_{m_1}\},\ldots,
          \{\rho^{k}_{m_{k}}\}}}^p
 \prod_{f'=1}^p\prod_{t'=1}^{l_t}\left[1+
 \delta_{f'\,s}\,\delta_{j_{f'}\,\varrho^t_{t'}}
 \left({a^{-1}(\m_{t},\l_\theta^{(1)})}
 -1\right)\right]
 \nonumber\\&& \quad\times
  \prod_{t=1}^k
 \prod_{\stackrel{s=1}{s\ne\{\rho^1_{m_1}\},\ldots,
          \{\rho^{k}_{m_{k}}\}}}^p
 \prod_{f'=1}^p \prod_{t'=1}^{l_t}\left[1+
 \delta_{f'\,s}\,\delta_{j_{f'}\,\varrho^t_{t'}}
 \left({a(\l_{s},\l_\theta^{(1)})}
 -1\right)\right]
 \nonumber\\ &&\mbox{}-
 \sum_{\e=k+1}^{\b-1}
 {b^{\mp}(\m_\b,\m_\e)\over a(\m_\b,\m_\e)}
 {b^pm(\l_\a,\m_\e)\over a(\l_\a,\m_\e)}
 \prod_{e=1}^k a^{l'_e}(\m_\e,\l_e)
 \prod_{t=1}^{k'}\prod_{t'=1}^{l'_t}
 {a^{-1}(\m_\e,\l_{\rho^t_{t'}})}
 \nonumber\\ &&\quad\times
  \prod_{\varepsilon=1,\ne \e}^{\b-1}
  {a^{-1}(\m_\e,\m_\varepsilon)}
 \prod_{\g=p+1}^{\a-1}{a^{-1}(\m_\e,\l_\g) }
  \left[\prod_{j=p+1}^{n_1}
  {a(\m_\e,\l_j^{(1)})}
 -\prod_{l=1}^N{a(\m_\e,\x_l)}\right]
   \nonumber\\ &&\quad\times
  \prod_{t=1}^k\
 \prod_{\stackrel{s=1}{s\ne\{\rho^1_{m_1}\},\ldots,
          \{\rho^{k}_{m_{k}}\}}}^p
 \prod_{f'=1}^p\prod_{t'=1}^{l_t}\left[1+
 \delta_{f'\,s}\,\delta_{j_{f'}\,\varrho^t_{t'}}
 \left({a(\m_\e,\m_{t})}
 -1\right)\right]
 \nonumber\\&& \quad\times
  \prod_{t=1}^k
 \prod_{\stackrel{s=1}{s\ne\{\rho^1_{m_1}\},\ldots,
          \{\rho^{k}_{m_{k}}\}}}^p
 \prod_{f'=1}^p \prod_{t'=1}^{l_t}\left[1+
 \delta_{f'\,s}\,\delta_{j_{f'}\,\varrho^t_{t'}}
 \left({a^{-1}(\m_\e,\l_{s})}
 -1\right)\right]
 \nonumber\\ &&\mbox{}+
 \sum_{e=1}^{k'}g^\pm_4(\m_\b,l'_e) +
 \sum_{t=1}^{k}(g')^\pm_4(\m_\b,l_t),
\end{eqnarray}
where $g_4^\pm(\m_\b,l'_e)$ is given as follows. i.) when
$\prod_{t=1}^{k'}\prod_{t'=1}^{l'_t}\delta_{e\, \rho^t_{t'}}=0$,
\begin{eqnarray}
 &&g_4^\pm(\m_\b,l'_e)=
 -\sum_{k=0}^{l'_e-1}{1\over k!}
 {e^{\mp\m_\b}\over
  \sinh(\m_\b-\l_e+\eta)^{l'_e-k}}
 {d^{k}\over d\chi^{k}_4}\left\{\sinh^{l'_e}(\m_\b-\l_e)
 \right. \nonumber\\ && \quad \times
 {e^{\pm\m_\b}b^{\pm}(\l_\a,\m_\b)\over a(\l_\a,\m_\b)}
 \prod_{e'=1,\ne e}^{k'}a^{l'_e}(\m_\b,\l_e)
 \prod_{t=1}^{k'}\prod_{t'=1}^{l'_t}
 {a^{-1}(\m_\b,\l_{\rho^t_{t'}})}
 \prod_{\g=p+1}^{\a-1}{a^{-1}(\m_\b,\l_\g)}
 \nonumber\\ &&\quad\times
 \prod_{\e=1}^{\b-1}{a^{-1}(\m_\b,\m_\e)}
  \left[\prod_{j=p+1}^{n_1}a(\m_\b,\l_j^{(1)})
 -\prod_{l=1}^N{a(\m_\b,\x_l)}\right]
\nonumber\\ &&\quad\times
  \prod_{t=1}^k
 \prod_{\stackrel{s=1}{s\ne\{\rho^1_{m_1}\},\ldots,
          \{\rho^{k}_{m_{k}}\}}}^p
 \prod_{f'=1}^p\prod_{t'=1}^{l_t}\left[1+
 \delta_{f'\,s}\,\delta_{j_{f'}\,\varrho^t_{t'}}
 \left({a(\m_\b,\m_{t})}
 -1\right)\right]
 \nonumber\\&& \quad\times \left.
  \prod_{t=1}^k
 \prod_{\stackrel{s=1}{s\ne\{\rho^1_{m_1}\},\ldots,
          \{\rho^{k}_{m_{k}}\}}}^p
 \prod_{f'=1}^p \prod_{t'=1}^{l_t}\left[1+
 \delta_{f'\,s}\,\delta_{j_{f'}\,\varrho^t_{t'}}
 \left({a^{-1}(\m_\b,\l_{s})}
 -1\right)\right] \right\}_{\chi_4=0}, \nonumber\\
\end{eqnarray}
and $\chi_4\equiv \chi_4(\m_\b)=\sinh(\m_\b-\l_e+\eta)$, ii.) when
$\prod_{t=1}^k\prod_{t'=1}^{l'_t}\delta_{e\, \rho^t_{t'}}=1$ and
$l'_e=1$, $g_4^\pm(\m_\b,l_e)=0 $ and  iii.) when there are
indices $\hat t$ ($\hat t\in \{1,\ldots k\}$) and $\hat t'$ ($\hat
t'\in \{1,\ldots l_{\hat t}\}$) such that $\rho^{\hat t}_{\hat
t'}=e$, and $l_e\geq 2$,
\begin{eqnarray}
 &&g^\pm_4(\m_\b,l_e)=
 -\sum_{k=0}^{l_e-2}{1\over k!}
 {e^{\mp\m_\b}\over
   \sinh(\m_\b-\l_e+\eta)^{l_e-k-1}}
 {d^{k}\over d\chi^{k}_4}\left\{\sinh^{l_e-1}(\m_\b-\l_e)
 \right. \nonumber\\ && \quad \times
  {e^{\pm\m_\b}b^{\pm}(\l_\a,\m_\b)\over a(\l_\a,\m_\b)}
 \prod_{e'=1,\ne e}^{k'}a^{l'_e}(\m_\b,\l_e)
 \prod_{t=1,\ne \hat t}^{k'}\,\prod_{t'=1,\ne \hat t'}^{l'_t}
 {a^{-1}(\m_\b,\l_{\rho^t_{t'}})}
  \nonumber\\ &&\quad\times
   \prod_{\g=p+1}^{\a-1}{a^{-1}(\m_\b,\l_\g)}
 \prod_{\e=1}^{\b-1}{a^{-1}(\m_\b,\m_\e)}
  \left[\prod_{j=p+1}^{n_1}{a(\m_\b,\l_j^{(1)})}
 -\prod_{l=1}^N{a(\m_\b,\x_l)}\right]
\nonumber\\ &&\quad\times
  \prod_{t=1}^k
 \prod_{\stackrel{s=1}{s\ne\{\rho^1_{m_1}\},\ldots,
          \{\rho^{k}_{m_{k}}\}}}^p
 \prod_{f'=1}^p\prod_{t'=1}^{l_t}\left[1+
 \delta_{f'\,s}\,\delta_{j_{f'}\,\varrho^t_{t'}}
 \left({a(\m_\b,\m_{t})}
 -1\right)\right]
 \nonumber\\&& \quad\times \left.
  \prod_{t=1}^k
 \prod_{\stackrel{s=1}{s\ne\{\rho^1_{m_1}\},\ldots,
          \{\rho^{k}_{m_{k}}\}}}^p
 \prod_{f'=1}^p \prod_{t'=1}^{l_t}\left[1+
 \delta_{f'\,s}\,\delta_{j_{f'}\,\varrho^t_{t'}}
 \left({a^{-1}(\m_\b,\l_{s})}
 -1\right)\right] \right\}_{\chi_4=0}; \nonumber\\
\end{eqnarray}
for the function $(g')^\pm_4(\m_\b,l_t)$, one has : i.)
$(g')^\pm_4(\m_\b,l_t)=0$ when
$$n_t\equiv \sum_{\stackrel{s=1}{s\ne\{\rho^1_{m_1}\},\ldots,
  \{\rho^{k}_{m_{k}}\}}}^p \sum_{f'=1}^p\sum_{t'=1}^{l_t}
 \delta_{f'\,s}\,\delta_{j_{f'}\,\varrho^t_{t'}}=1,$$
ii.) when $n_t=0$,
\begin{eqnarray}
&&(g')^\pm_4(\m_\b,l_t)=
 -{b^\mp(\m_\b,\m_t)\over a(\m_\b,\m_t)}
  {b^\pm(\l_\a-\m_t)\over a(\l_\a,\m_t}
 \prod_{e=1}^k a^{l'_e}(\m_t,\l_e)
 \prod_{t=1}^{k'}\prod_{t'=1}^{l'_t}
 {a^{-1}(\m_t,\l_{\rho^t_{t'}})}
 \nonumber\\ &&\quad\times
  \prod_{\e=1,\ne t}^{\b-1}{a^{-1}(\m_t,\m_\e)}
 \prod_{\g=p+1}^{\a-1}{a^{-1}(\m_t,\l_\g)}
  \left[\prod_{j=p+1}^{n_1}
  {a(\m_t,\l_j^{(1)})}
 -\prod_{l=1}^N{a(\m_t,\x_l)}\right]
   \nonumber\\ &&\quad\times
  \prod_{\tau=1,\ne t}^k
 \prod_{\stackrel{s=1}{s\ne\{\rho^1_{m_1}\},\ldots,
          \{\rho^{k}_{m_{k}}\}}}^p
 \prod_{f'=1}^p\prod_{\tau'=1}^{l_\tau}\left[1+
 \delta_{f'\,s}\,\delta_{j_{f'}\,\varrho^\tau_{\tau'}}
 \left({a(\m_t,\m_{\tau})}
 -1\right)\right]
 \nonumber\\&& \quad\times
  \prod_{\tau=1}^k
 \prod_{\stackrel{s=1}{s\ne\{\rho^1_{m_1}\},\ldots,
          \{\rho^{k}_{m_{k}}\}}}^p
 \prod_{f'=1}^p \prod_{\tau'=1}^{l_\tau}\left[1+
 \delta_{f'\,s}\,\delta_{j_{f'}\,\varrho^\tau_{\tau'}}
 \left({a^{-1}(\m_t,\l_{s})}
 -1\right)\right], 
\end{eqnarray}
iii.) when $n_t\geq 2$,
\begin{eqnarray}
&&(g')^\pm_4(\m_\b,n_t)=-\sum_{k=0}^{n_t-2}{1\over k!}
 {e^{\mp\m_\b}\over \sinh(\m_\b-\m_e+\eta)^{n_t-k-1}}
 {d^{k}\over d(\chi'_4)^{k}}
 \left\{\sinh^{-1}(\m_t-\m_\b)
 \right.
 \nonumber\\&&\quad \times
 \prod_{\stackrel{s=1}{s\ne\{\rho^1_{m_1}\},\ldots,
          \{\rho^{k}_{m_{k}}\}}}^p
 \prod_{f'=1}^p\prod_{t'=1}^{l_t}
 \left[\sinh(\m_{t}-\m_\b -\eta)
 \right.\nonumber\\ &&\quad\quad\quad\mbox{}\left.+
 \delta_{f'\,s}\,\delta_{j_{f'}\,\varrho^t_{t'}}
 (\sinh(\m_{t}-\m_\b)-\sinh(\m_{t}-\m_\b -\eta))
 \right]
 \nonumber\\&&\quad \times
   {e^{\pm\m_\b}b^{\pm}(\l_\a,\m_\b)\over a(\l_\a,\m_\b)}
 \prod_{e=1}^k a^{l'_e}(\m_\b,\l_e)
 \prod_{t=1}^{k'}\prod_{t'=1}^{l'_t}
 {a^{-1}(\m_\b,\l_{\rho^t_{t'}})}
 \prod_{\g=p+1}^{\a-1}{a^{-1}(\m_\b,\l_\g)}
 \nonumber\\ &&\quad\times
 \prod_{\e=1,\ne t}^{\b-1}{a^{-1}(\m_\b,\m_\e)}
  \left[\prod_{j=p+1}^{n_1}{a(\m_\b,\l_j^{(1)})}
 -\prod_{l=1}^N{a(\m_\b,\x_l)}\right]
 \nonumber\\ &&\quad\times
  \prod_{\tau=1,\ne t}^k
 \prod_{\stackrel{s=1}{s\ne\{\rho^1_{m_1}\},\ldots,
          \{\rho^{k}_{m_{k}}\}}}^p
 \prod_{f'=1}^p\prod_{\tau'=1}^{l_\tau}
 \left[1+
 \delta_{f'\,s}\,\delta_{j_{f'}\,\varrho^\tau_{\tau'}}
 \left({a(\m_\b,\m_{\tau})}-1\right)\right]
 \nonumber\\&& \quad\times \left.
  \prod_{\tau=1}^k
 \prod_{\stackrel{s=1}{s\ne\{\rho^1_{m_1}\},\ldots,
          \{\rho^{k}_{m_{k}}\}}}^p
 \prod_{f'=1}^p \prod_{\tau'=1}^{l_\tau}
 \left[1+
 \delta_{f'\,s}\,\delta_{j_{f'}\,\varrho^\tau_{\tau'}}
 \left({a^{-1}(\m_\b,\l_{s})}-1\right)\right]
  \right\}_{\chi'_4=0}, \nonumber\\
\end{eqnarray}
and $\chi'_4\equiv \chi_2(\m_\b)=\sinh(\m_\b-\m_e+\eta)$.

Therefore from (\ref{eq:Sn-Gn}) and
(\ref{eq:Gp-G1G2})-(\ref{de:Gm-Gm-1234}), we have the following
theorem:
\begin{Theorem}
Let the spectral parameters $\{\l_k\}$ of the Bethe state
$|\O_N(\{\l_k\}_{(p,n_1)})\rangle $ be solutions of the $BAE$
(\ref{eq:BAE}). The scalar products
$\mathbb{P}_{n_1}(\{\m_k\}_{(p,n_1)},\{\l_k\}_{(p,n_1)})$ defined
by (\ref{de:P_{n_1}}) are represented by
\begin{eqnarray}
 && \mathbb{P}_{n_1}(\{\m_k\}_{(p,n_1)},\{\l_k\}_{(p,n_1)})
 \nonumber\\
 &&=(-1)^{n_1}\sum_{\s,\s'\in{\cal S}_{n_1}}
 Y_L(\{\m_{\sigma'(j)}\},\{\m^{(1)}_{\sigma'(k)}\})
 Y_R(\{\l_{\sigma(j)}\},\{\l^{(1)}_{\sigma(k)}\})\,
 \nonumber\\&& \quad \times \left\{
 \mbox{det}\, {\cal F}^-(\l_{\s(1)},\ldots,\l_{\s(p)};
         \m_{\s'(1)},\ldots,\m_{\s'(p)})
 \right.
 \nonumber\\ && \quad\quad\quad \times
 \mbox{det}\,{\cal N}^-_1(\l_{\s(p+1)},\ldots,\l_{\s(n_1)};
      \m_{{\s'(p+1)}},\ldots,\m_{{\s'(n_1)}})
 \nonumber\\ &&\quad\quad \mbox +
 {\sum_{k,l_k,\varrho^k_{m_k}}}'
 \sum_{\stackrel{j_1,\ldots,j_p=1}{j_1\ne\ldots\ne j_p}}^p
 (-1)^{\tau(\s'(j_1)\ldots \s'(j_p))}
 (A_2^-)_{\s(1)\,\s'(j_1)}\ldots (A_2^-)_{\s(p)\,\s'(j_p)}
 \nonumber\\ && \quad\quad\quad\times
 \mbox{det}\,{\cal N}^-_2(\l_{\s(p+1)},\ldots,\l_{\s(n_1)};
      \m_{{\s'(p+1)}},\ldots,\m_{{\s'(n_1)}};\{l_k\})
 \nonumber\\ &&\quad\quad \mbox +
 {\sum_{k,l_k,\rho^k_{m_k}}}''
 \mbox{det}\, A_3^-(\l_{\s(1)},\ldots,\l_{\s(p)};
         \m_{\s'(1)},\ldots,\m_{\s'(p)})
 \nonumber\\ &&\quad\quad\quad \times
 \mbox{det}\,{\cal N}_3^-(\l_{\s(p+1)},\ldots,\l_{\s(n_1)};
      \m_{{\s'(p+1)}},\ldots,\m_{{\s'(n_1)}};\{l_k\})
 \nonumber\\ &&\quad\quad  \mbox +
 {\sum_{k,l_k,\varrho^k_{m_k}}}'
  {\sum_{k',l'_{k'},\rho^{k'}_{m'_{k'}}}}''
 \sum_{\stackrel{j_1,\ldots,j_p=1}{j_1\ne\ldots\ne j_p}}^p
 (-1)^{\tau(\s'(j_1)\ldots \s'(j_p))}
 \nonumber\\ &&\quad \quad\quad\times
 (A_4^-)_{\s(1)\,\s'(j_1)}\ldots (A_2^-)_{\s(p)\,\s'(j_p)}
 \nonumber\\ &&\quad \quad\quad\times \left.
 \mbox{det}\,{\cal N}_4^-(\l_{\s(p+1)},\ldots,\l_{\s(n_1)};
      \m_{{\s'(p+1)}},\ldots,\m_{{\s'(n_1)}};\{l_k\};\{l'_{k'}\})\right\}.
    \no\\  \label{eq:theorem}
\end{eqnarray}
\end{Theorem}


{\bf Remark:} In the derivation of (\ref{eq:theorem}), the
spectral parameters $\{\l_i\}$ in the state
$|\O_N(\{\l_j\}_{(p,n_1)})\rangle$ are required to satisfy the BAE
(\ref{eq:BAE}). However, the parameters $\m_j$ $(j=1,\ldots,n_1)$
in the dual state $\langle\O_N(\{\m_j\}_{(p,n_1)})|$ do not need
to satisfy the BAE.

On the other hand, if we compute the scalar product by starting
from the dual state $\langle\O_N(\{\l_j\}_{(p,n_1)})|$, then by
using the same procedure, we have
\begin{eqnarray}
 &&\mathbb{P}_{n_1}^L(\{\l_k\}_{(p,n_1)},\{\m_j\}_{(p,n_1)})\no\\
 &=&(-1)^{n_1}\sum_{\s,\s'\in{\cal S}_{n_1}}Y_L(\{\l_{\s(j)}\},\{\l^{(1)}_{\s(k)}\})
 Y_R(\{\m_{\s'(j)}\,\{\m^{(1)}_{\s'(k)}\})\,\nonumber\\ && \times
 G_L^{(n_1)}(\{\m_{\s'(j)}\}_{(p,n_1)},\{\l_{\s(k)}\}_{(p,n_1)}),
 \label{eq:Pn-L}
\end{eqnarray}
where the function $G_L$, compared with $G$ in
(\ref{de:Gm-Gm-1234}), is given by $G_L\equiv\sum_{i=1}^4
(G^+)_i$. In (\ref{eq:Pn-L}), we have also assumed that any
element of the spectral parameter set $\{\l_i\}$ satisfy the BAE.


\subsection{Determinant representation of correlation  functions}

Having obtained the scalar product and the norm, we are now in the
position to compute the k-point correlation functions of the
model. In general, a $l$-point correlation function of the local
generators $E^{i,j}_{(k)}$ is defined by
\begin{eqnarray}
\langle E^{i_1,j_1}_{(k_1)}\ldots E^{i_l,j_l}_{(k_l)}\rangle
 \equiv\langle\O_N(\{\m_j\})|E^{i_1,j_1}_{(k_1)}\ldots E^{i_l,j_l}_{(k_l)}
  |\O_N(\{\l_j\})\rangle. \label{de:cf-general}
\end{eqnarray}

In principle, by using the theorem 2 and theorem 6, we may compute
any correlation function defined by (\ref{de:cf-general}). As an
example, in this subsection, we compute the correlation function
associated with two adjacent generators $E^{3,2}_{(\k)}$ and
$E^{2,3}_{(\k+1)}$.
\begin{Proposition}
If both the Bethe state $|\O_N(\{\m_j\})\rangle$ and the dual
Bethe state $\langle\O_N(\{\m_j\})|$ ($\m_j=\l_j,\m_j$) are
eigenstates of the transfer matrix, then the two-point correlation
functions associated with the local generators $E^{3,2}_{(\k)}$
and $E^{2,3}_{(\k+1)}$ can be represented by
\begin{eqnarray}
 &&\langle\O_N(\{\m_j\})_{(p,n_1)}|E^{3,2}_{(\k)}E^{2,3}_{(\k+l)}
  |\O_N(\{\l_j\})_{(p,n_1)}\rangle
 \nonumber\\ &&
 =\sum_{\s\in{\cal S}_{n_1}}\sum_{\s'\in{\cal S}_{n_1}}
 \phi_{\kappa-1}(\{\m_{j}\})
 \phi^{-1}_{\kappa+1}(\{\l_{k}\})
 \sum_{i=1}^p (-1)^{i-1}
  {b^+(\x_{\k+1},\l_{\s(i)})\over a(\x_{\k+1},\l_{\s(i)})}
 \nonumber\\ && \quad\times \left\{
 \sum_{j=i+1}^p{b^+(\x_{\k+1},\l_{\s(j)})\over a(\x_{\k+1},\l_{\s(j)})}
 \prod_{l=i+1}^{j-1}
  {c(\l_{\s(l)},\l_{\s(j)})\over a(\l_{\s(l)},\x_{\k})}
 \prod_{k=i+1,\ne j}^{p}
  {c(\l_{\s(j)},\l_{\s(k)})\over a(\l_{\s(j)},\l_{\s(k)})}
  \right.
 \nonumber\\ && \quad\quad\times
 \prod_{\a=1}^N a(\l_{\s(j)},\x_\a)\,
 {\cal K}\left(i+1;\l_{\s(i)};\{\m_{\s'(d)}\}_{(p,n_1)};
  \{\l'_{\s(f)}\}_{(p,n_1)}\right)
 \nonumber\\ && \quad\quad \mbox{}
 -\prod_{j=i+1}^p
  {c(\l_{\s(i)},\l_{\s(j)})\over a(\l_{\s(i)},\l_{\s(j)})}
 \prod_{\a=1}^N a(\l_{\s(i)},\x_\a)\,
 \no\\ && \quad\quad\times
 {\cal K}\left(i+1;\x_{\k+1};\{\m_{\s'(d)}\}_{(p,n_1)};\{\l^*_{\s(f)}\}_{(p,n_1)}\right)
 \nonumber\\ && \quad\quad\mbox{}
 -\sum_{j=i+1}^p{b^+(\l_{\s(i)},\l_{\s(j)})\over a(\l_{\s(i)},\l_{\s(j)})}
 \prod_{l=i+1}^{j-1}
  {c(\l_{\s(l)},\l_{\s(j)})\over a(\l_{\s(l)},\l_{\s(i)})}
 \prod_{k=i+1,\ne j}^{p}
  {c(\l_{\s(j)},\l_{\s(k)})\over a(\l_{\s(j)},\l_{\s(k)})}
 \nonumber\\ && \quad\quad\, \times \left.
 \prod_{\a=1}^N a(\l_{\s(j)},\x_\a)
 {\cal K}\left(i+1;\x_{\k+1};\{\m_{\s'(d)}\}_{(p,n_1)};
    \{\l''_{\s(f)}\}_{(p,n_1)}
  \right)\right\},
\end{eqnarray}
where $\phi_i(\{\m_j\})=\prod_{k=1}^i\prod_{l=1}^n
a^{-1}(\m_l,\x_k)$, ${\cal K}$ is given by
\begin{eqnarray}
 &&{\cal K}(e;\delta;\{\m_{\s'(j)}\};\{\l_{\s(k)}\}) \nonumber\\
 &&= \prod_{i=e}^{n_1} {1\over a(\l_{\s(i)},\delta)}
 P^L_{n_1}\left(\{\m_{\s'(j)}\}_{(p,n_1)};\{\l_{\s(k)}\}_{(p,n_1)}\right)
 \nonumber\\ && \quad\mbox{}
 -\sum_{j=p+1}^{n_1}
  {b^-(\l_{\s(j)},\delta)\over a(\l_{\s(j)}, \delta)}
 \prod_{k=p+1}^{j-1}
  {c(\l_{\s(k)},\l_{\s(j)})\over c( \l_{\s(k)},\delta)}
 \prod_{l=e,\ne j}^{n_1} {1\over a(\l_{\s(l)},\l_{\s(j)})}
 \nonumber\\ && \quad\,\times
 P^L_{n_1}\left(\{\m_{\s'(d)}\}_{(p,n_1)};\{\l_{\s(1)},\ldots,\l_{\s(e)},\ldots,
  \l_{\s(j-1)},\delta,\l_{\s(j+1)},\ldots,\l_{\s(n_1)}\}_{(p,n_1)}\right)
 \nonumber\\ && \quad\mbox{}
 -\sum_{i=e}^p{b^-(\l_{\s(i)},\delta)\over a(\l_{\s(i)},\delta)}
  \prod_{k=1}^{i-1}
   {c(\l_{\s(k)},\l_{\s(i)})\over c( \l_{\s(k)},\delta)}
 \prod_{j=1,\ne i}^{p} {1\over a(\l_{\s(j)},\l_{\s(i)})}
 \left[
 \prod_{l=p+1}^{n_1} {1\over a(\l_{\s(l)},\l_{\s(i)})}\right.
 \nonumber\\ && \quad\,\times
 P^L_{n_1}\left(\{\m_{\s'(d)}\}_{(p,n_1)};\{\l_{\s(1)},\ldots,\l_{\s(e)},\ldots,
  \l_{\s(i-1)},\delta,\l_{\s(i+1)},\ldots,\l_{\s(n_1)}\}_{(p,n_1)}\right)
 \nonumber\\ && \quad\mbox{}
 -\sum_{l=p+1}^{n_1}
  {b^-(\l_{\s(l)},\l_{\s(i)})\over a(\l_{\s(l)},\l_{\s(i)})}
 \prod_{m=p+1}^{l-1}
  {c(\l_{\s(m)},\l_{\s(l)})\over c( \l_{\s(m)},\l_{\s(i)})}
 \prod_{q=p+1,\ne l}^{n_1} {1\over a(\l_{\s(q)},\l_{\s(l)})}
 \nonumber\\ && \quad\,\times 
 P^L_{n_1}\left(\{\m_{\s'(d)}\}_{(p,n_1)};\{\l_{\s(1)},\ldots,\l_{\s(e)},\ldots,
  \l_{\s(i-1)},\delta,\l_{\s(i+1)},\ldots, \right.
 \nonumber\\ && \quad\quad\quad\quad\quad\quad\left.\left.
  \l_{\s(l-1)},\l_{\s(i)},\l_{\s(l+1),\ldots,\l_{\s(n_1)}}\}_{(p,n_1)}
  \right)\right]
  \label{de:P-cal}
\end{eqnarray}
and the spectral parameters $\l',\l^*$ and $\l''$ are given by
\begin{eqnarray*}
 &&\l'_{\s(k)}=\left\{\begin{array}{cl}
 \x_\k & \quad\quad(k=1)\\
 \l_{\s(k-1)} &\quad\quad(2\leq k\leq i)\\
 \l_{\s(k)} & \quad\quad(i+1\leq k\leq n_1 \mbox{ and } k\ne j)\\
 \x_{\k+1} &\quad\quad(k=j)\end{array} \right.,
 \\ &&
 \l^*_{\s(k)}=\left\{\begin{array}{cl}
 \x_\k & \quad\quad(k=1)\\
 \l_{\s(k-1)} &\quad\quad(k=2,\ldots,i)\\
 \l_{\s(k)} & \quad\quad(i+1\leq k\leq n_1)\end{array} \right.,\quad
 \\ &&
 \l''_{\s(k)}=\left\{\begin{array}{cl}
 \x_\k & \quad\quad(k=1)\\
 \l_{\s(k-1)} &\quad\quad(2\leq k\leq i)\\
 \l_{\s(k)} &\quad\quad (i+1\leq k\leq n_1 \mbox{ and } k\ne j)\\
 \l_{\s(i)} &\quad\quad (k=j)\end{array} \right.,
\end{eqnarray*}
respectively.
\end{Proposition}

\noindent {\it Proof.} From the definition (\ref{de:cf-general}),
the correlation function is written by
\begin{eqnarray}
 &&\langle \O^p_N(\{\m_j\})_{(p,n_1)}| E^{3,2}_{(\k)}E^{2,3}_{(\k+1)}|
   \O^p_N(\{\l_k\})_{(p,n_1)}\rangle
 \nonumber\\ &
 =&\langle \O^p_N(\{\m_j\})_{(p,n_1)}| \prod_{j=1}^{\k-1}t(\x_j)
  C_2(\x_\k)B_2(\x_{\k+1}) \prod_{j=\k+2}^{N}t(\x_j)
 |\O^p_N(\{\l_k\})_{(p,n_1)}\rangle \label{eq:cf-proof-1}\\
&=&\prod_{j=1}^{n_1}\prod_{k=1}^{\k-1}a^{-1}(\m_{j},\x_k)
 \prod_{j=1}^{n_1+1}\prod_{k=\k+1}^{N}a^{-1}(\l_{j},\x_k)\no\\
&&\times\langle \O^p_N(\{\m_j\})_{(p,n_1)}|
  C_2(\x_\k)\,B_2(\x_{\k+1})
 |\O^p_N(\{\l_k\})_{(p,n_1)}\rangle\label{eq:cf-proof-2} \\
  &=& \sum_{\s\in {\cal S}_{n_1}}\sum_{\s'\in{\cal S}_{n_1}}
 \phi_{\kappa-1}(\{\l_{\s(j)}\})
 \phi^{-1}_{\kappa+1}(\{\m_{\s'(k)}\})
    Y_L(\{\m_{\s'(j)}\},\{\m_{\s'(k)}^{(1)}\}) \no\\ &&\times
    Y_R(\{\l_{\s(j)}\},\{\l_{\s(k)}^{(1)}\})
    \langle0|\overleftarrow{\prod_{i=p+1}^{n_1}}B_1(\m_{\s'(i)})
    \overleftarrow{\prod_{i=1}^{p}}B_2(\m_{\s'(i)})
  \nonumber\\ && \times
    C_2(\x_\k)B_2(\x_{\k+1})\,
    \overrightarrow{\prod_{i=1}^{p}} C_2(\l_{\s(i)})
    \overrightarrow{\prod_{i=p+1}^{n_1}}
    C_1(\l_{\s(i)})|0\rangle,\label{eq:cf-proof-3}
\end{eqnarray}
where in (\ref{eq:cf-proof-1}), we have used the theorem 2 and the
property: for the $q$-deformed supersymmetric $t$-$J$ model with
periodic boundary condition, the transfer matrices satisfy the
relation $\prod_{i=1}^N t(\l_i)=1$; in (\ref{eq:cf-proof-2}), we
have used the theorem 5; and (\ref{eq:cf-proof-3}), we have used
the relation $ \prod_{j=1}^{n}\prod_{k=1}^Na^{-1}(\l_j,\x_k)=1,$
which is from the BAE and the NBAE.

Then with the help of the following commutation relations
\begin{eqnarray}
 && C_a(\l)C_a(\m)=-c(\l,\m)C_a(\m)C_a(\l),
 \label{eq:commu-CaCa}\\
 &&A_{ab}(\l)C_c(\m)=
 {r(\l-\m)^{bc}_{de}\over a(\l-\m)}C_e(\m)A_{ad}(\l)
 +{b^+(\l-\m)\over a(\l-\m)}C_b(\l)A_{ac}(\m),\label{eq:commu-AC}\\
 &&D(\l)C_c(\m)={1\over a(\m,\l)} C_c(\m)D(\l)-
 {b^-(\m,\l)\over a(\m,\l)} C_c(\l)D(\m), \label{eq:commu-DC}\\
 &&B_a(\l)C_b(\m)=-C_b(\m)B_a(\l)+{b^+(\l,\m)\over a(\l,\m)}[
 D(\m)A_{ab}(\l)-D(\l)A_{ab}(\m)]
\end{eqnarray}
and the theorem 6, we arrive at this proposition.
~~~~~~~~~~~~~~~~~~~~~~~~~~~~~~~~~~~~~~~~~~~~~~~~~~~~~~$\Box$

\sect{Conclusion and outlook}

We have reviewed our recent progress on the construction of the
determinant representations of the correlation functions for
supersymmetric fermion models via the algebraic Bethe ansatz. The
main idea was to simplify the creation (or annihilation) operators
and therefore the Bethe state (or its dual state) with the help of
the Drinfeld twists. In the $F$-basis, the creation operators and
the Bethe states can be represented in completely symmetric forms.
This leads to the scalar products of Bethe states represented by
determinants. The determinant representations of the correlation
functions were then constructed by means of the scalar products.
The determinant representations are useful for analysing
asymptotics of time and temperature dependant correlation
functions \cite{Kitanine02}${}^-$\cite{Gromov05} which are
important in statistical mechanics and condensed matter physics.
They also have applications in algebraic combinatorics and
alternating sign matrices.

It would be interesting to generalize the construction of the
correlation functions to other integrable models, e.g. the
elliptic $Z_N$ Belavin model (for which the symmetric
representations of the Bethe states was obtained in
\cite{Albert0007}), the supersymmetric $U$ model \cite{yzz95}, the
EKS model \cite{Essler9211} and the Hubbard model.

\vspace{4mm}


{\bf Acknowledgements:} This work was financially supported by the
Australian Research Council. S.Y. Zhao was supported by the UQ
Postdoctoral Research Fellowship.


\end{document}